\documentclass[aps,prd,twocolumn,showpacs,preprintnumbers,nofootinbib,letterpaper]{revtex4}

\usepackage{color}
\usepackage{graphicx}
\usepackage{eepic}
\usepackage{times}
\usepackage{subfigure}
\usepackage{amsmath}
\usepackage{amssymb}
\usepackage{fancyhdr}
\usepackage{hyperref}

\DeclareGraphicsExtensions{.pdf}

\begin{document}
\rhead[]{}
\lhead[]{}

\title{Efficient generation and optimization of stochastic template banks by a neighboring cell algorithm}

\author{Henning Fehrmann}
\email{henning.fehrmann@aei.mpg.de}
\author{Holger~J.~Pletsch}
\email{holger.pletsch@aei.mpg.de}
\affiliation{
 Max-Planck-Institut f\"ur Gravitationsphysik (Albert-Einstein-Institut),
 D-30167 Hannover, Germany, and\\
 Institut f\"ur Gravitationsphysik, Leibniz Universit\"at Hannover,
 D-30167 Hannover, Germany
}
\date{\today}


\begin{abstract}
Placing signal templates (grid points) as efficiently as possible to cover a multi-dimensional 
parameter space is crucial in computing-intensive matched-filtering searches for 
gravitational waves, but also in similar searches in other fields of astronomy. 
To generate efficient coverings of arbitrary parameter spaces, stochastic template banks 
have been advocated, where templates are placed at random while
rejecting those too close to others. However, in this simple scheme, for each new 
random point its distance to \emph{every} template in the existing bank is computed.
This rapidly increasing number of distance computations can render 
the acceptance of new templates computationally prohibitive,
particularly for wide parameter spaces or in large dimensions.
This work presents a  \emph{neighboring cell algorithm} that can
dramatically improve the efficiency of constructing a stochastic template bank.
By dividing the parameter space into sub-volumes (cells),
for an arbitrary point an efficient hashing technique is exploited to
obtain the index of its enclosing cell along with the parameters of its neighboring templates.
Hence only distances to these neighboring templates in the bank are
computed, massively lowering the overall computing cost, as demonstrated 
in simple examples.
Furthermore, we propose a novel method based on this technique to increase the 
fraction of covered parameter space solely by directed template shifts, without adding any 
templates. As is demonstrated in examples, this method can be highly effective.
\end{abstract}

\maketitle

\section{introduction}

In searches for gravitational-wave signals using matched-filtering methods
or similar detection statistics, efficient template banks play an essential role
\cite{PhysRevD.44.3819,PhysRevD.49.1707,PhysRevD.49.2658,lrr-2005-3,Babak+2012,arXiv:1207.7176}
when analyzing data from the
ground-based detector instruments such as LIGO \cite{Abbott_2004,Abramovici_1992},
Virgo \cite{Acernese_2006}, GEO600 \cite{Gossler2002,Willke2002}
and TAMA \cite{Takahashi_2004}, as well as future space-based
detectors \cite{Danzmann1998}.
Searches for signals based on template banks are also relevant in neighboring
research fields of astronomy, for example in binary pulsar searches of radio data
from the Arecibo radio telescope \cite{Benk2010,Allen+2013} or gamma-ray data
from the \emph{Fermi} satellite \cite{Pletsch+2012-J1311}.

In the standard procedure of matched-filtering searches, the instrumental
data is correlated with a {\it template} that has the form of the expected signal.
Because the parameters of the signal are unknown a priori, the data must be
correlated with a {\it bank} (or grid) of possible signal templates (grid points)
that have distinct parameter values \cite{Helstrom1968,Whalen1971}.
In particular for wide search parameter spaces,
methods for constructing template banks that minimize the 
computational burden without decreasing the signal detectability
are essential. This can be achieved by placing the templates more optimally,
such that fewer are required at the same level of signal detectability.
Different strategies for improved template placement have been studied in previous work, 
e.g. see \cite{Manca_2010} and references therein.

For Euclidean spaces, the problem of finding an optimal \emph{lattice} coverings is 
well studied \cite[e.g.,][]{Conway1993} and is related to periodic crystalline structures.
For up to 5 dimensions, the so-called $A_{n}^{*}$ lattice turns out to cover the space
with the least number of templates \cite{magdeburg,Vallentin2003}. 
For more than 5 dimensions, other lattice structures
are also known which have a better covering behavior than $A_{n}^{*}$
lattices \cite{magdeburg}. However, a crystalline structure is often not optimal 
if either a covering fraction of less than one is desired, or the parameter space 
is not Euclidean or has large dimensionality.

To address such cases, alternative schemes for arbitrary parameter spaces 
have been considered by Messenger et al. \cite{messenger2008},
where templates are simply placed at random.
Such {\it random} template banks show superior covering behavior in higher dimensions
compared to relaxed (less than 100\% covering fraction) $A_{n}^{*}$ lattices.
In low dimensions, Manca and Vallisneri~\cite{Manca_2010}  showed that the covering of random template 
banks can be improved by exploiting Sobol quasi random
sequences \cite{sobol_1967,sobol_1976,lampert_1988}.
However, with an increasing number of dimensions
the resulting improvement becomes less pronounced.

In {\it stochastic} template banks that have been suggested by
Harry, Allen and Sathyaprakash~\cite{harry2009} (hereafter HAS09) as well as by Babak~\cite{babak_2008} 
(hereafter B08), templates are picked at random too, 
but only those are added to the bank which have a distance larger
than a certain predefined value (covering radius) to any of those templates already 
in the bank. This procedure is to continue until no more new template can be added to the bank.
Compared to (fully) random template placement, the filtering stage done for a stochastic template bank
thus results in a more diluted template bank, leading in turn to a more efficient search 
since much fewer templates have to be evaluated.
As HAS09 note, for this approach the number of templates to reach 
a covering fraction of one (i.e. full coverage) is actually finite.

However, a major drawback of the existing stochastic template bank algorithms
is the severe computational complexity involved in the
bank construction. This is mainly because the distance between each new candidate template 
and \emph{every} other template already part of the bank has to be computed and 
compared to the covering radius before acceptance.
Especially for covering fractions approaching one, the computational cost 
for accepting a new template to the bank increases much more rapidly 
than quadratically, and can in fact become prohibitive
for large or high-dimensional parameter spaces.

In this work, we present a solution to this problem
by dividing the parameter space into smaller sub-volumes (cells)
to drastically reduce the aforementioned computational burden. This basic idea is hardly new,
but inspired by previous work \citep[][]{PhysRevLett.64.119, PhysRevD.79.022001}, 
and also alluded to in~HAS09. Here we develop an efficient concept, which we refer to 
as the neighboring cell algorithm (NCA), and study its performance improvements. 
In addition, we describe a new method to significantly increase the covering 
fraction of a stochastic template bank \emph{without} having to add further templates to bank,
but performing systematic shifts of the templates in the bank.
 
This paper is organized as follows. To set the stage, Sec.~\ref{s:StochasticBanks}
describes the standard stochastic template bank generation algorithm
along with some general properties of stochastic banks. 
Section~\ref{s:nca} presents the NCA algorithm to efficiently generate stochastic 
template banks and also demonstrates its computational performance. 
Building on this, in Sec.~\ref{s:shifts} we show how the NCA can also be 
exploited to optimize a stochastic template bank by purely 
shifting the existing templates of a bank, leading to a significantly increased 
covering fraction without additional templates. Section~\ref{s:examples}
illustrates two example applications for different parameter spaces. 
In Sec.~\ref{s:generalhca}, we describe a scheme to further generalize the NCA 
to arbitrarily complicated parameter spaces by combining cells 
to adaptive virtual cells. Finally, this is followed by a conclusion and
discussion of future directions in Sec.~\ref{s:conclusion}.

\section{Stochastic Template Placement}
\label{s:StochasticBanks}

To set the notation, let $M$ label the $d$-dimensional 
signal manifold, which we refer to as parameter space. We also assume the
availability of a positive-definite distance function for two points in $M$. 
Thus, a template bank consisting
of $N$ points in parameter space is said to completely cover $M$ with covering radius~$r$
if every point in $M$ lies within a distance $r$ of at least one of the $N$ 
points of the template bank. An optimal template bank covering $M$
with radius $r$ would be most economical, i.e. having the minimum number of points.

The standard algorithm for stochastic template placement that has been proposed 
in HAS09 (and B08) uses the following principal scheme, 
beginning with an empty template bank:
\begin{enumerate}
\item[(1)] 
Draw a random point in $M$ and add it to the template bank.
\item[(2)]
Draw another random point in $M$ and add it to the template bank only 
if its distance is greater than $r$ to \emph{every} 
other already accepted point of the template bank. 
\item[(3)]
Repeat the previous step until
the number of points
in template bank stops changing, or other termination criteria are fulfilled.
\end{enumerate}

In general, the distance between two points in $M$ (or two normalized signals) 
measures their overlap (i.e. \emph{match}). Another way to think of it, is that one
of the two points is a template and the other is a signal, so that the distance
reflects the fractional loss in signal-to-noise ratio (SNR) due to the parameter
offsets (i.e. \emph{mismatch}) between the template and the signal.

The key difference between the methods of HAS09 and B08
is the distance computation between points. 
Whereas B08 uses the exact overlap to compute the distances,  
HAS09 exploit a computationally less expensive approximation
to the overlap via the geometric concept of a metric on parameter 
space \cite{Sathy1:1996,owen:1996me}. This metric tensor is obtained by 
Taylor-expanding the fractional loss in squared SNR to quadratic order
in the parameter offsets. Hence, for problems where a reliable analytic metric is 
available across $M$, the cost of distance calculations can be significantly reduced.
Additional efficiency can be achieved in such cases by also 
modulating the distribution of random candidate templates according
to the volume element given by the metric, as discussed in HAS09.

However, whether or not using a metric approximation is used, in either case 
the total number of required distance comparisons in step~(2) of this basic scheme 
can quickly render the entire construction process computationally intractable. 
The NCA proposed here presents a solution to this problem to significantly 
reduce the number of distance computations needed, as will be described 
in Sec.~\ref{ss:Performance}.

To assess the efficiency of a template covering, typically its {\it thickness} $\Theta$
or its  {\it normalized thickness} $\theta$ are considered.
We follow \cite{Conway1993,messenger2008,harry2009} and refer to the thickness $\Theta$ as
the average number of templates covering any parameter-space point. 
The normalized thickness is just \mbox{$\theta =\Theta / V_d$},
where \mbox{$V_d$} is the volume of a $d$-dimensional unit sphere,
\mbox{$V_d = \pi^{-d/2} \Gamma(\frac{d}{2}+1)$}.
Both, $\Theta$ and $\theta$ are invariant properties of the covering, independent
of $r$.  The total number of templates, $N = \theta \, r^{-d} \, V_M$, is thus directly 
proportional to the normalized thickness (where $V_M$ is the proper volume of $M$).

\begin{figure}
\centering
 \includegraphics[width=0.47\textwidth]{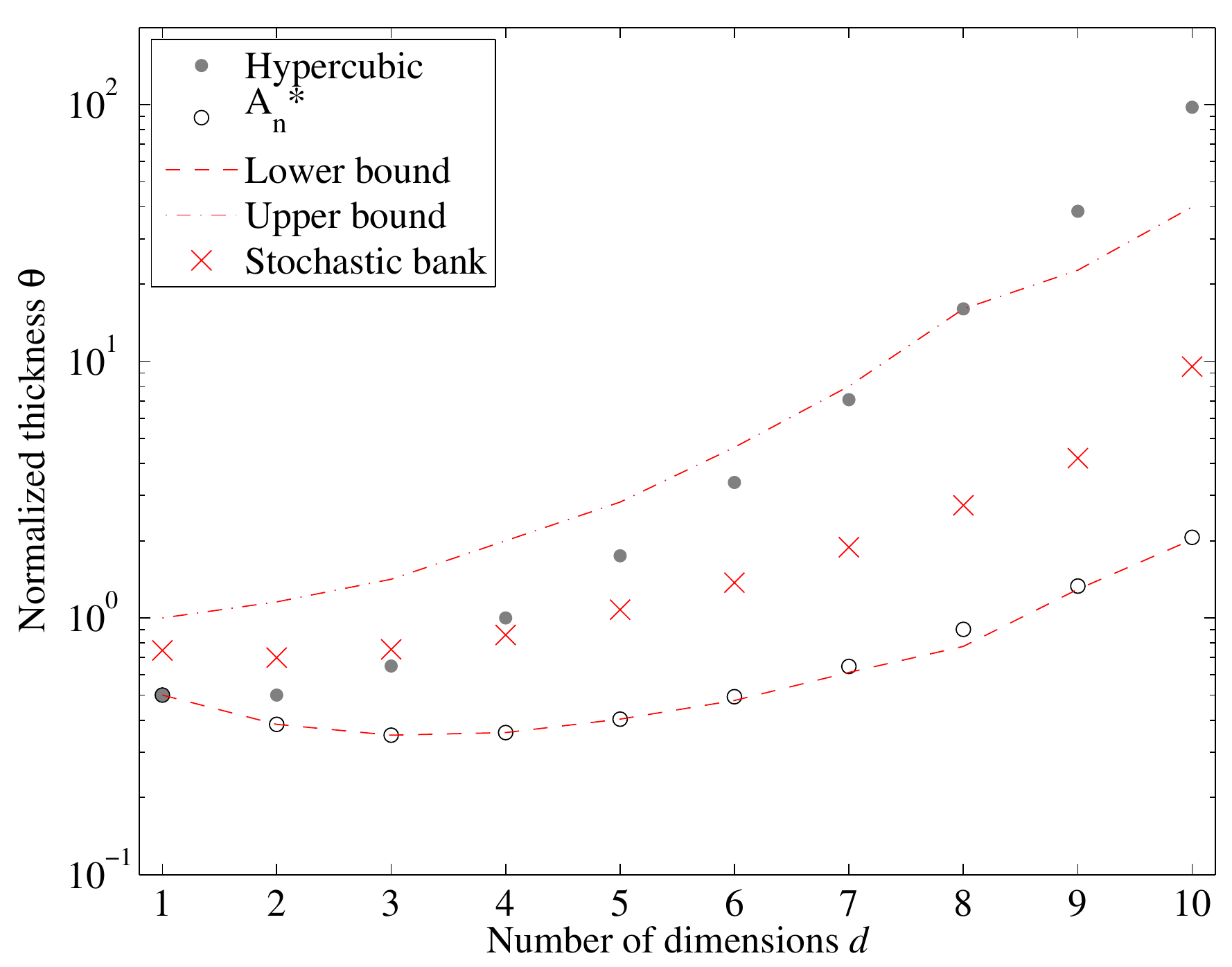}
 \caption{
  \label{fig:normalizedthickness}
  The normalized thickness~$\theta$ as a function of the number of
  dimensions~$d$ for a stochastic template bank in $\mathbb{E}^d$
  (crosses).
  In addition, the lower and upper bounds for stochastic template banks
  are shown, derived from the known minimum and maximum possible 
  normalized thickness values for the sphere packing problem
  \cite{magdeburg,Conway1993} as described in the text. 
  For comparison, also lattice placement algorithms (hyper-cubical and $A_n^*$) 
  are displayed.
}
\end{figure}

As outlined in HAS09, one can obtain theoretical upper and lower bounds 
for the required number of templates for a stochastic bank with complete covering.
A theoretical upper bound follows from the sphere packing problem by considering
how many non-overlapping spheres with radius $r/2$ can be packed into a certain
volume. As the center of such hard spheres are separated by $r$, 
these are possible positions for a stochastic template bank. Thus, we use the
results of the sphere packing problem given in \cite{Conway1993} 
to obtain the upper bounds on the normalized thickness 
shown in Fig.~\ref{fig:normalizedthickness} 
A theoretical lower bound on the number of required
templates for a complete coverage is $V_M/(V_d\, r^d)$, which is the ratio of the
parameter-space volume and the volume of one template with radius $r$.
Therefore, the best possible thickness is $\Theta=1$, i.e. the best possible
normalized thickness is $\theta=1/V_d$.
In practice however it is impossible to reach this for $d>1$ \cite{Conway1993}.

For Euclidean spaces~$\mathbb{E}^d$ with $d$ dimensions a large body of literature
exists \cite{Conway1993} seeking to find the lattice with minimum
possible thickness, such that when placing a $d$-dimensional sphere
with radius~$r$ at each lattice point, the set of all such spheres completely covers $\mathbb{E}^d$.
Figure~\ref{fig:normalizedthickness} shows the known minimum 
possible normalized thickness values as listed in~\cite{magdeburg}.
For comparison, the performance of  two other lattice algorithms is displayed,
the hyper-cubical lattice and the so-called $A_n^*$ lattice \cite{Conway1993}.
Notice that a hyper-cubical lattice in dimensions $d>4$ for unit covering fraction 
does not satisfy the conditions of a stochastic template bank.

Extending Fig.~1 of HAS09, we are also able to compute the
normalized thickness for a stochastic template bank for up to 10 dimensions.
This calculation was only made possible owing to the significantly improved 
computational efficiency of the NCA algorithm presented in this work. 
Due to computational limitations, in HAS09 the results for only up to 4 
dimensions have been reported. For dimensions higher than 3, the stochastic bank
increasingly improves in performance compared to the simple hyper-cubical lattice.
Though the stochastic bank with complete coverage appears to be less efficient than
the $A_n^*$ lattice.  However, any such lattice coverings and constructions algorithms 
are only defined for the cases of flat parameter spaces. In contrast, the stochastic template
bank generation with the NCA in principle can be done for any parameter space and
thus can prove especially useful for non-flat parameter spaces.
 
When generating a stochastic template bank, 
it is useful to define the {\it covering fraction} $f \in [0,1]$, 
which denotes the faction of $V_M$ that is covered by the union 
of all template volumes. As HAS09 note, $f$ is also related to the
computational complexity to populate the
stochastic template bank, because the expected number of
candidate templates required to accept one template to the bank 
increases as $1/(1-f)$.

\section{The neighboring cell algorithm}
\label{s:nca}

\subsection{Key elements and requirements of the NCA}
\label{ss:keyelements}

In the following we describe the key elements and basic
requirements of the NCA in order to efficiently generate  a stochastic 
template bank:
\begin{enumerate}
	\item[(1)] The entire parameter space is divided into non-overlapping
	  sub-volumes that we denote as \emph{cells}.
	  For simplicity and ease of computing, we employ hyper-cubical
	  cells in a Cartesian coordinate system.\footnote{
	  For maximum efficiency in curved (i.e. non-flat) parameter spaces,
	  it its recommended to adapt the cells sizes to follow the local metric
	  approximation. See Sec.~\ref{s:examples} for an example.
	  For complicated curved parameter spaces (e.g. where widely separated 
	  cells would have high overlap), a possible alternative 
	  is outlined in Sec.~\ref{s:generalhca} using the concept of virtual cells.
	  }
	  Any other regular or non-regular structure is also possible.
	\item[(2)] Each cell must be uniquely indexed.
	\item[(3)] Generally, two cells are neighbors if at least one common parameter-space point
	  of one cell and a template covering volume exists where the
	  template lies inside the other cell. In Cartesian coordinates
	  and hyper-cubical cells neighboring cells have at least one
	  border point in common.
	  For a given cell index, the indices of the neighboring
	  cells can be computed.
	\item[(4)] Each template must be uniquely indexed.
	\item[(5)] Each template index is mapped to a cell index that labels
	  the cell in which the template position lies.
	\item[(6)] Given a parameter-space position of a template, the index of its
	  enclosing cell is readily computed by a hashing algorithm.
	  For a hyper-cubical cell lattice this is
	  accomplished by a fast rounding or truncating operation on the
	  parameter values of the template position.
	\item[(7)] \label{size_of_template} The parameter-space  
	  covered by a template only overlaps with the space of the enclosing
	  cell and the neighboring cells.  It must not overlap with
	  the space of non-neighboring cells. This is ensured by
	  either choosing a small enough covering radius of the templates,
	  or by designing the cells sufficiently large.
\end{enumerate}

\begin{figure}
  \begin{center}
   \includegraphics[width=75mm,bb=55 120 400 200]{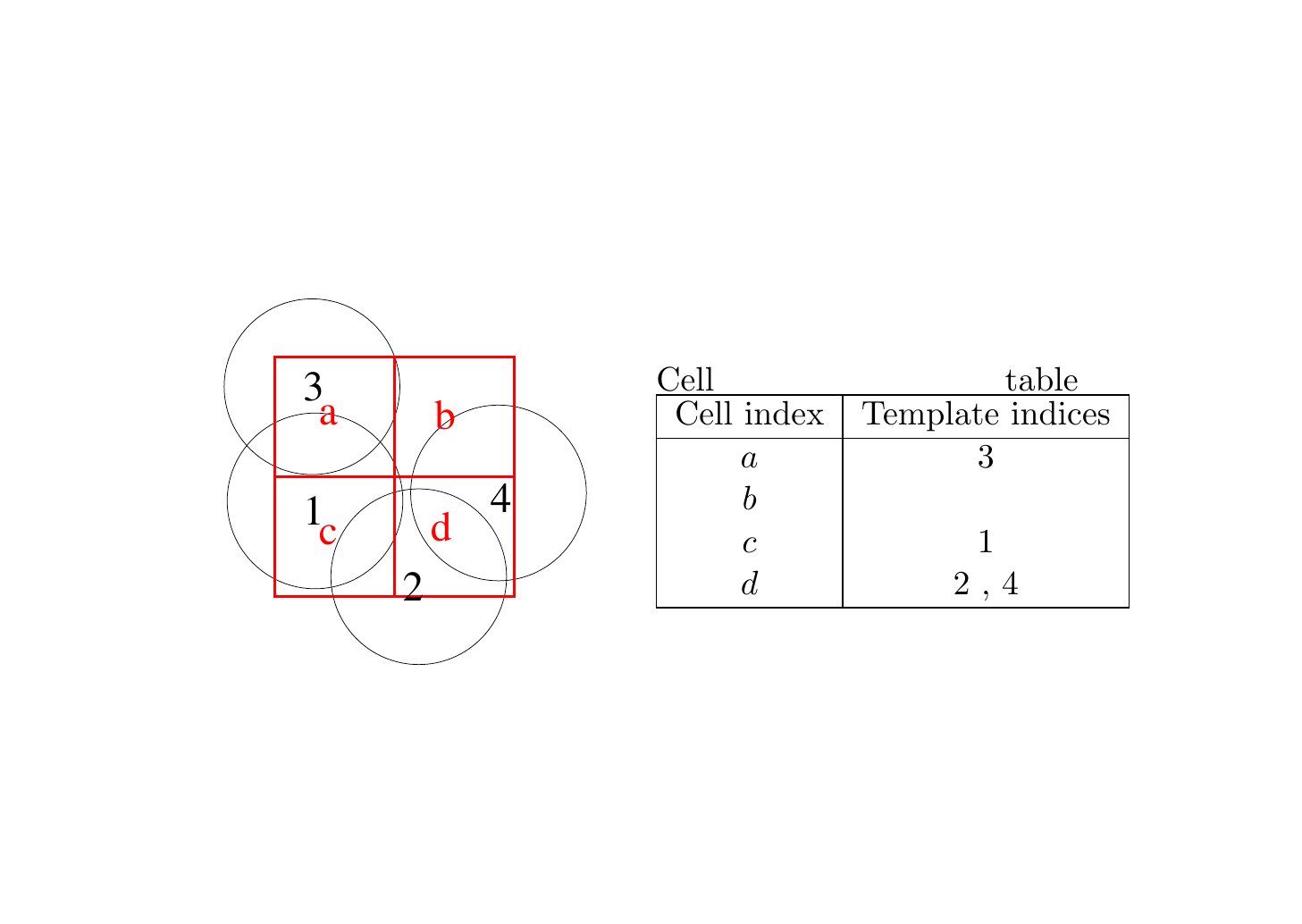}
  \end{center}
\caption{ \label{fig:hash_cell}
Exemplary illustration of the cell table,
which contains all cell indices  and the list of template indices of template 
positions lying within each cell. 
 Shown on the left are four templates (black circles) labeled by indices~$\{1,2,3,4\}$.
The cells (red boxes) are labeled by the indices~$\{a,b,c,d\}$.
}
\end{figure}

The NCA requires two different tables to be kept in memory. 
First, the \emph{template table}, which stores the indices and 
the parameter-space positions of all templates.
Second, the \emph{cell table}, which contains all cell indices 
and the list of template indices of template positions lying within each cell.
Figure~\ref{fig:hash_cell} shows a simplified 
example for the contents of the cell table.

\subsection{Stochastic template placement with the NCA}
\label{ss:stobankgeneration}

\begin{figure}
  \begin{center}
    \includegraphics[width=75mm,bb=65 100 350 200]{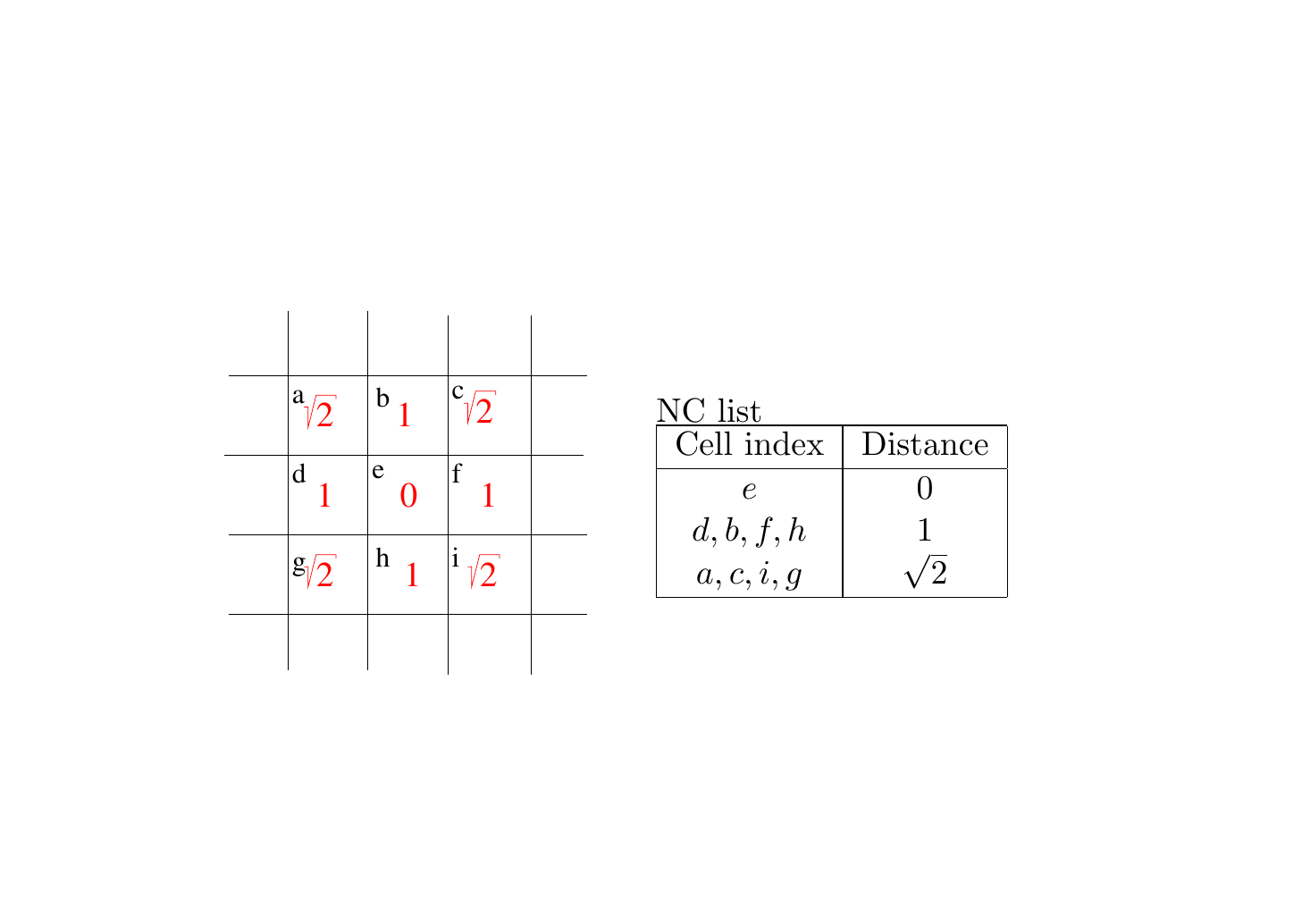}
  \end{center}
\caption{
  \label{fig:cellorder}
  Schematic illustration of the distances between neighboring cells of the NC list.
  The letters $\{a,b,c,d,e,f,g,h,i\}$ represent the cell indices. The numbers
  shown are the distances from the center of the cell labeled by~$e$
  to each center of the other cells shown. 
  For maximum efficiency, the NC list is sorted
  by increasing order of this distance
  as explained in the text.
}
\end{figure}

The following procedure describes how to generate a stochastic template
bank with covering fraction~$f$ using the NCA. 
Starting with an empty template bank, the process for adding 
new templates to the bank is as follows:
\begin{enumerate}
	\item[(1)]
	Draw a random\footnote{A pseudo-, quasi or real random number generator may be used.
	As described in HAS09, when an analytic metric is available across $M$, it is advisable
	to modulate the distribution of random points according
        to the volume element given by the metric for maximum efficiency.} 
	point in $M$ as a candidate template for addition to the bank.	
	\item[(2)]
	Determine the cell the candidate template falls into by
	calculating its cell index $i_{\rm C}$.
	\item[(3)]
	Compute the cell indices of all neighboring cells. We refer to the resulting list
	of neighboring cells including the enclosing cell $i_{\rm C}$ as the
	\emph{neighboring-cell} (NC) list. Note that a cell that is a $d$-dimensional hypercube, 
	 has ($3^d-1$) neighboring cells.
        \item[(4)] \label{i:cell_sorting}
	Sort the NC list in order of increasing distance.
	Start with $i_{\rm C}$ and proceed with the closest neighboring cells.
	The distance between two cells is defined as the distance between 
	the centers of the cells. For illustrative purposes, Fig.~\ref{fig:cellorder} 
	shows an example for the NC list.
	\item[(5)]
	Retrieve a list of all template indices associated with the cells
	of the NC list.
	\item[(6)]
	For all retrieved template indices obtain the
	template positions from the template table.
	\item[(7)]
 	Compute the distance between the candidate template and
	every other template position of those obtained in the previous step.
	Start with the templates located in cells nearest to cell $i_{\rm C}$.
	If all computed distances exceed the predefined covering radius,
	accept this candidate template as a new template and 
	assign it the next consecutive template index.
	\item[(8)]
	If the candidate template has been accepted, update the template table:
	Append the position of the accepted candidate template to the
	template table.
	Also update the cell table: Append the template index to the list of template 
	indices pertaining to the cell $i_{\rm C}$ in the cell table.
	\item[(9)]
	Repeat this procedure starting from step~(1) until the covering
	fraction~$f$ has reached the desired value.
\end{enumerate}

\subsection{Computational cost improvements from the NCA}
\label{ss:Performance}

A major drawback of the standard stochastic template bank algorithm by HAS09 
is the computational complexity which can become even prohibitive. 
This is because for each new candidate template its distance to every
other template of the existing bank has to be computed and compared
to the covering radius before eventual acceptance to the bank.
With increasing covering fraction the probability of
accepting a candidate template decreases. At the beginning, when the 
covering fraction substantially less than one, the estimated computational cost 
for accepting a new candidate template increases approximately 
quadratically with the number of templates in the bank,
since almost no templates are rejected. For covering fractions closer 
to one the computational cost increases much faster than quadratically
with the number of templates in the bank, because
the rejection of candidate templates dominates.
It is this prohibitive computing cost that can quickly render the acceptance
of new templates computationally intractable in the standard 
stochastic template bank algorithm.

\begin{figure}[t]
  \hspace{-10mm}
  \includegraphics[width=0.95\linewidth]{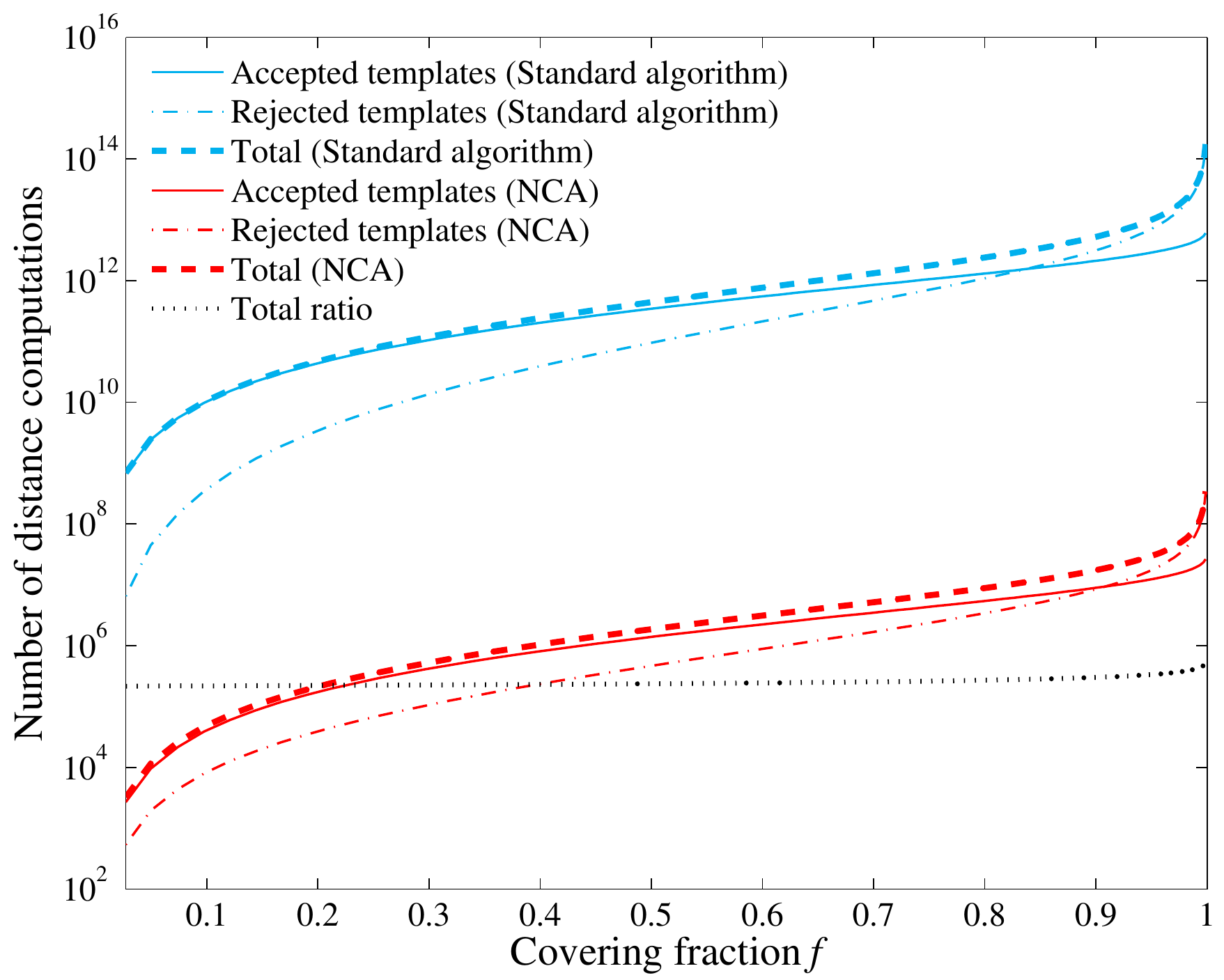}
  \caption{Comparison the computing cost (measured by the number of distance computations)
  between the standard algorithm and the NCA to generate a stochastic template bank in $\mathbb{E}^3$.
  Shown as a function of the covering fraction~$f$  for each method is
  the number of distance computations for candidate templates that
  have been accepted (thin solid) and rejected (thin dotted-dashed), as
  well as the total number (thick dashed).
  With the NCA, the total number of distance computations is massively reduced,  
  by more than five orders of magnitude as shown by the dotted curve. 
  \label{fig:speedup} 
}
\end{figure}

In order to demonstrate the significant  computational efficiency improvement  
of the NCA over the standard algorithm, we consider the generation of 
a stochastic template bank in a 3-dimensional Euclidean space with periodic boundary 
conditions, where each coordinate lies within $[0,1]$. The template covering
radius $r$ is chosen, such that the bank contains $3.5$ million templates 
at a covering fraction of $f=99.9\%$, which is given by $r=1/180$ and 
implies a normalized thickness of about $0.6$.
To compare the computational costs of both algorithms when constructing this bank,
we count the total number of distance computations required in either case. 
Figure~\ref{fig:speedup} shows the results of this comparison study. With the NCA, the total 
number of distance computations is massively reduced and about five orders of magnitude
lower compared to the standard algorithm. 
This gain factor is not surprising but straightforward to understand: It is simply
the ratio of the total volume of the considered parameter space to the volume
enclosed by a single cell and its $(3^3-1)$ neighboring cells, which gives
$180^3/27 \approx 2\times 10^5$ in this example.

Figure~\ref{fig:speedup} also shows the shares in the total number of distance
computations separately for rejected and accepted candidate templates.
As can be seen, at the beginning for low covering fractions, the computing
cost is dominated by the distance computations for accepted candidate templates.
As the template bank is getting more populated at higher covering fractions,
a turnover occurs, where  the computing cost starts being dominated 
by the distance computations due to rejections and increases
much more rapidly. As can be seen, compared to the standard algorithm 
for the NCA this turnover takes place at a larger covering fraction.
This is mainly due to the NCA's much more efficient rejection of candidate
templates, as described in the following.

Sorting the NC list of neighboring cells is crucial for the efficiency of the NCA,
specifically in view rejecting candidate templates.
This sorting [done in step~(4) in Sec.~\ref{ss:Performance} above] as part of 
the NCA stochastic template bank generation  
considerably reduces the average number of distance computations needed
before a candidate template is eventually rejected.  
As illustrated in Fig.~\ref{fig:cellorder}, this is obvious, because
on average the overlapping volume of the candidate template is highest with
the {\it own} cell and decreases for the neighboring cells.
Therefore, the probability for rejecting a candidate template is the highest
when comparing to those templates located in the same cell.  Hence, the
sorting of the NC list by distance of step~(4) can also be seen as sorting by 
decreasing order of probability of rejection, which thus overall 
minimizes the average number of distance computations. This is also
seen in that the gain factor between the NCA and the standard scheme 
(dotted curve in Fig.~\ref{fig:speedup}) is mostly constant
but increases for covering fractions closer to one, where the rejections of
candidate templates dominate.

With higher dimensions this effect gains even more importance, because
the number of neighboring cells increases exponentially with dimension.
For example, if $d=3$ the number of considered cells is
$3^{3}=27$.  Without sorting and a cell addressing scheme as shown in
Fig.~\ref{fig:cellorder}, the cell containing the candidate would be on
average the $(3^{3}+1)/2 =14$th cell considered.  After sorting, the
cell enclosing the template candidate is considered first.  Hence, this
sorting can decrease the number of distance computations by a factor
almost $14$. Whereas, if $d=10$ the number of considered 
cells is $3^{10}=59049$.  The cell containing the candidate in absence
of sorting would be the $(3^{10}+1)/2 =29525$th cell considered.
Therefore, sorting the NC list can decrease the number of distance
computations in $d=10$ by almost $30\,000$! This efficiency gain has
greatest importance for covering fractions nearing one, where the
majority of candidate templates is rejected (see
Fig.~\ref{fig:speedup}).

The NCA also significantly facilitates evaluating the covering fraction
at the different stages of the stochastic bank generation. The covering fraction is 
typically obtained via Monte Carlo integration using a sufficient number of sample 
points (as also done in HAS09). The standard algorithm by HAS09 has to compute the distances
between a sample point and all templates in the bank, which is inefficient. 
The NCA instead readily can provide a list of the subset of templates closest to 
a given sample point, and only the distances to those are computed. 
This way, wasteful distance computations for templates far away from the sample-point 
location are avoided, as those templates will obviously have no overlap with the 
sample point.

\section{Increasing the covering fraction by shifting templates}
\label{s:shifts}

The generation of stochastic template banks with covering fractions 
nearing unity can become quickly computationally prohibitive (cf. Fig.~\ref{fig:speedup}).
This is due to the enormous number of candidate templates to be tested before a new 
template is accepted to the bank. Here, we present a possible and efficient
alternative solution to this problem. The idea is to first generate a stochastic 
template bank with initially smaller covering fraction
and then increase the covered space by only {\it shifting} the positions of 
the templates, instead of adding new ones.

\subsection{Barycentric template shifts}

In what follows, we describe a scheme to effectively shift the templates
in the bank with the goal of increasing the overall covering fraction.
One such shift optimization stage begins with the first template in bank:
\begin{enumerate}
	\item[(1)] 
	  Determine\footnote{One way to do this is described in detail in Sec.~\ref{ss:boundarypoints}.} a set of points uniformly distributed on the boundary of 
	  the covering volume of the template. Note that the \emph{boundary} 
	  of a covering volume is the set of points which have distance $r$ 
	  (the covering radius) to the template position.
	\item[(2)] 
	  Check whether each of these points is covered or not by another 
	  template.\footnote{Notice that this
can be accomplished by treating each boundary point like
candidate template as described in Sec.~\ref{s:nca}.} 
	  If covered the boundary point gets the zero weight,
	  otherwise a weight of unity. In case a boundary point lies outside 
	  of the relevant parameter space this point gets also zero weight.
	 \item[(3)]  
	 From the set of boundary points with unit weight, 
          calculate the barycenter of these points.
	\item[(4)] 
	  If the distance between the template position and the barycenter is 
	  smaller then a certain maximum distance~$\epsilon$, the template 
	  is moved to coincide with the barycenter.
	  If the distance is larger than $\epsilon$, the template position moved 
	  in the direction of the barycenter by only~$\epsilon$.
	\item[(5)]
	  Carry out the procedure starting from step~(1) for the next template
	  until done for all templates in the bank.
\end{enumerate}
The above scheme (forming a single optimization stage) is to be repeated
until the covering fraction does not increase anymore (or any other
terminating condition is met). 
In general, step~(4) will increase the fraction of covered parameter space.
However, it might also happen that occasionally a template is shifted towards 
an existing template, leading to an undesired newly created overlap.  
To mitigate this effect, we therefore recommend to set the maximum shift 
distance~$\epsilon$ at a fraction of $r$.
Within this work we found that choosing a maximum shift of $\epsilon = 0.05\,r$
provides overall satisfactory results. 

\subsection{Choice of boundary points and computing cost}\label{ss:boundarypoints}

The actual number of boundary points used is a tradeoff between 
accuracy of the barycentric shift and computational efficiency.
As a lower bound, to be able to shift the template position 
into any direction in a $d$-dimensional parameter space, 
the minimum number of points should is $2 d$.
More boundary points will improve the accuracy of the shift, but
also decrease the computational efficiency of a single optimization stage.
While a more detailed study of these aspects is beyond
the scope of this paper, one scheme we found to work sufficiently well
for our purposes is choosing twice as many boundary points as there are
neighboring cells, i.e. $2\times(3^d-1)$ points.

\begin{figure}
  \begin{center}
  \subfigure
    {\includegraphics[width=0.45\linewidth]{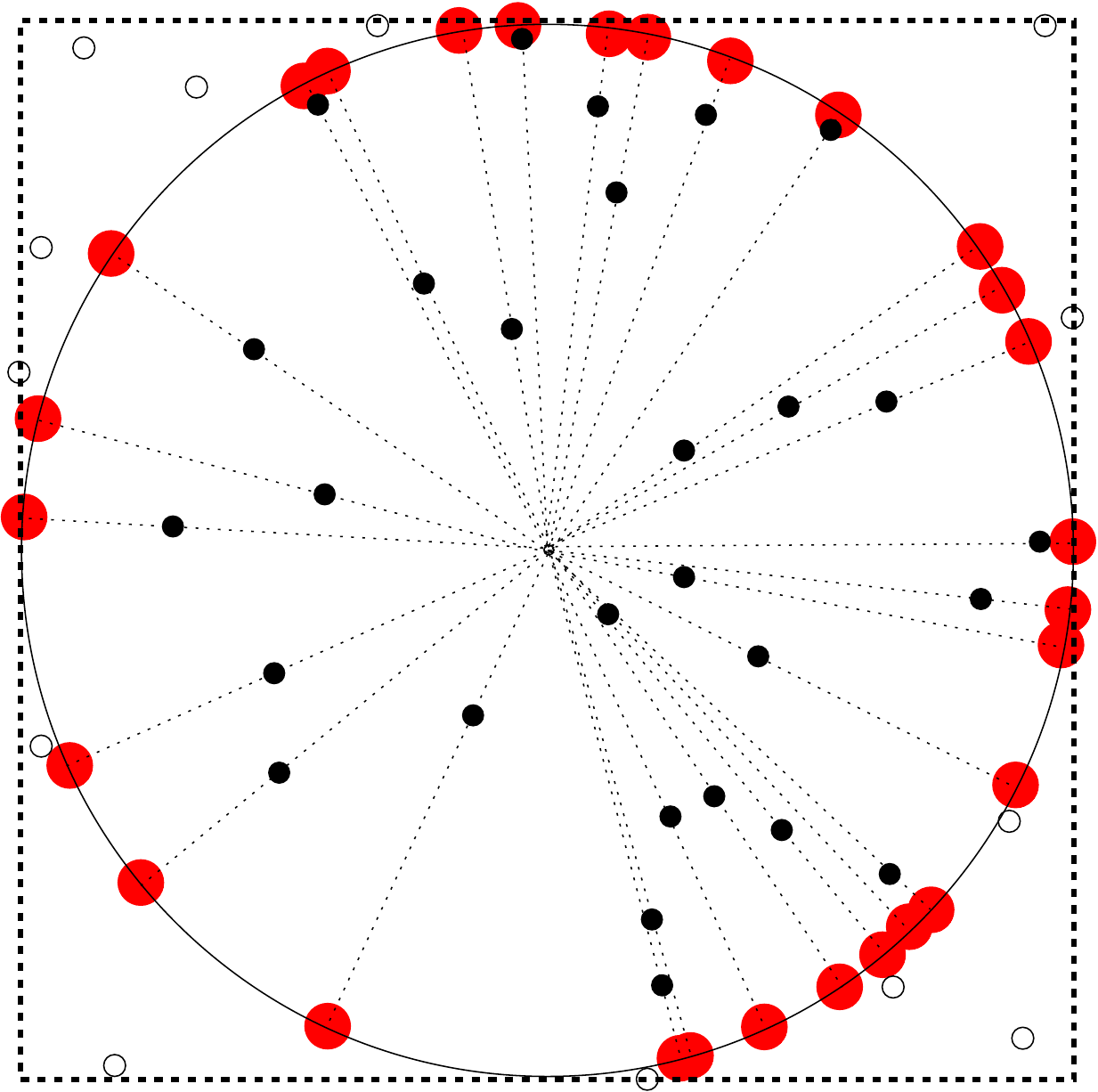}}
    \hspace{5mm}
    \subfigure
    {\includegraphics[width=0.38\linewidth]{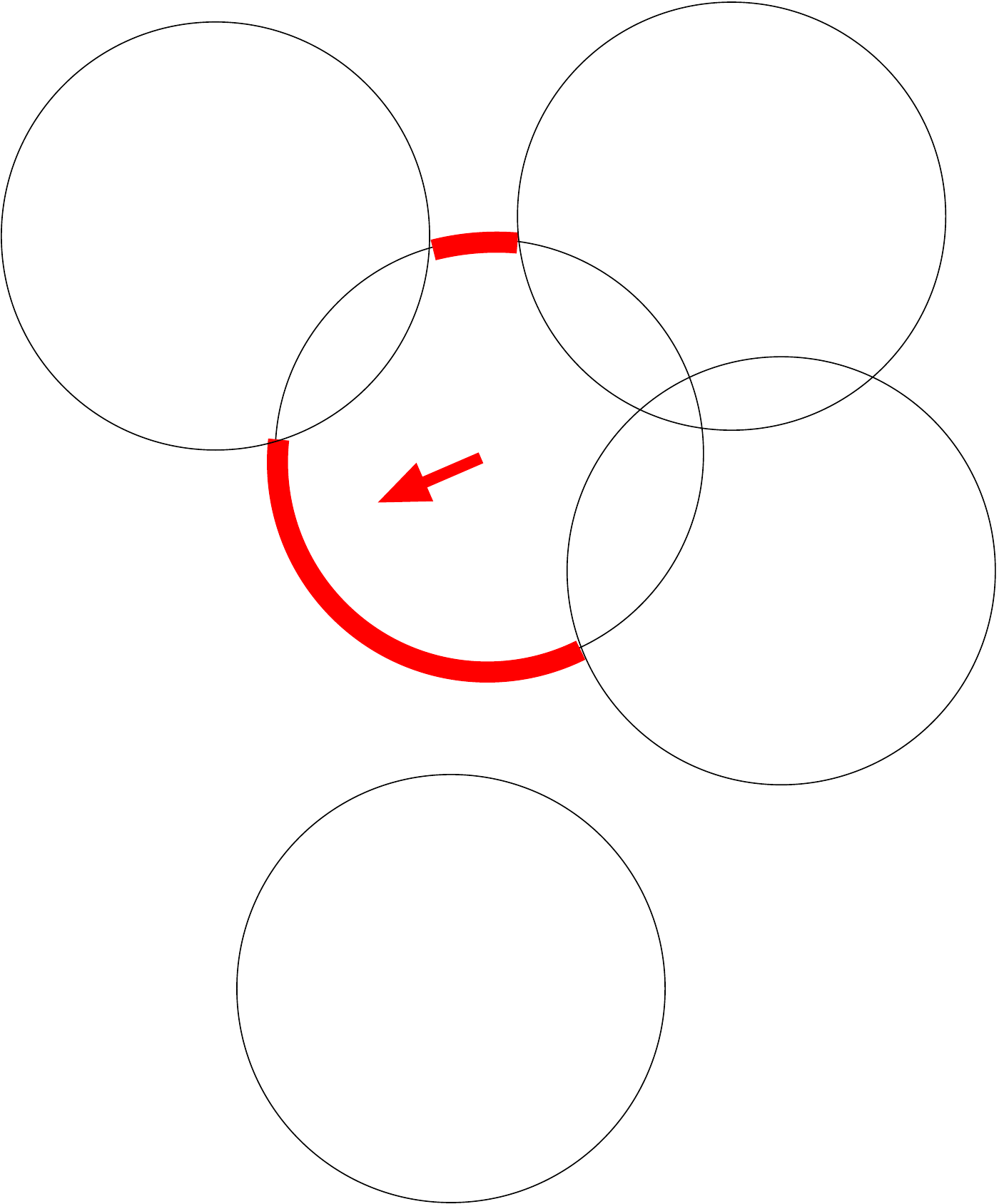}}
   \end{center}
\caption{
\label{fig:optimizing_scheme}
Schematic illustration of the barycentric template shifts 
to increase the covering fraction.
Left: Illustration of the generation of the boundary points. In the
enclosing cube around the sphere random points are placed
uniformly. The points falling outside of the template volume (small
hollow points) are ignored. The points are inside the sphere 
(small filled black points) they are projected onto the boundary 
of the template volume, providing the boundary points.
Right: The thick red lines shows points on the boundary of a
template covering volume that have weight one. These points are 
not covered by any neighboring template.  All other boundary points 
get the weight zero. Using these weights we can compute the barycenter 
of the boundary points. The arrow points from the current position of the 
template to the center of mass of the unit-weight (red thick) boundary points.  
The resulting shift of the template position towards 
this barycenter thus increases the covering fraction.
}
\end{figure}

One way to place the set of boundary points as required in step~(1)  
is the following approach, first presented in \cite{neumann1951,cook1957}.
In a $d$-dimensional Euclidean space with spherical 
template volumes, a uniformly distributed set of boundary points 
can be obtained by placing random points uniformly into the enveloping
hyper-cubical box. 
As illustrated in Fig.~\ref{fig:optimizing_scheme}, then all points 
lying outside of the sphere are discarded and  those inside the sphere are 
projected onto the boundary to provide the desired set of boundary points.

The computational cost of this optimization scheme 
is again dominated  by the number of distance
computations. In this method, the number of needed
distance computations~$D$ is simply
the product $D = N_O \times N_B \times N_S \times N \times $, where
$N_O$ is the number of optimization stages,
$N_B$ is the number of boundary points,
$N_S$ is the average number of templates in the considered within the subvolume of the neighboring cells,
and $N$ is the total number of templates.
As mentioned above, $N_B$ can be taken as  $2\times(3^d-1)$. Moreover,
$N_S$ is estimated as the normalized thickness $\theta$ times
the number of neighboring cells plus the own cell giving $3^d$.
The value of $N_O$ depends on the used shifting method, while
typically we reached convergence after $N_O=20$ stages.
Thus, for example in three dimensions, the number of required distance
computations using the normalized thickness of the optimal
covering is \mbox{$D = 20\times26\times 9.8\times N \approx 10200\times N$}.
It is worth noting that since the
computational cost is only linear in the total number of templates $N$,
this proposed scheme is feasible also for a relatively large template banks.

\begin{figure*}
\hspace{-5mm}
   \subfigure[]
  {\includegraphics[width=0.3\linewidth,angle=0]{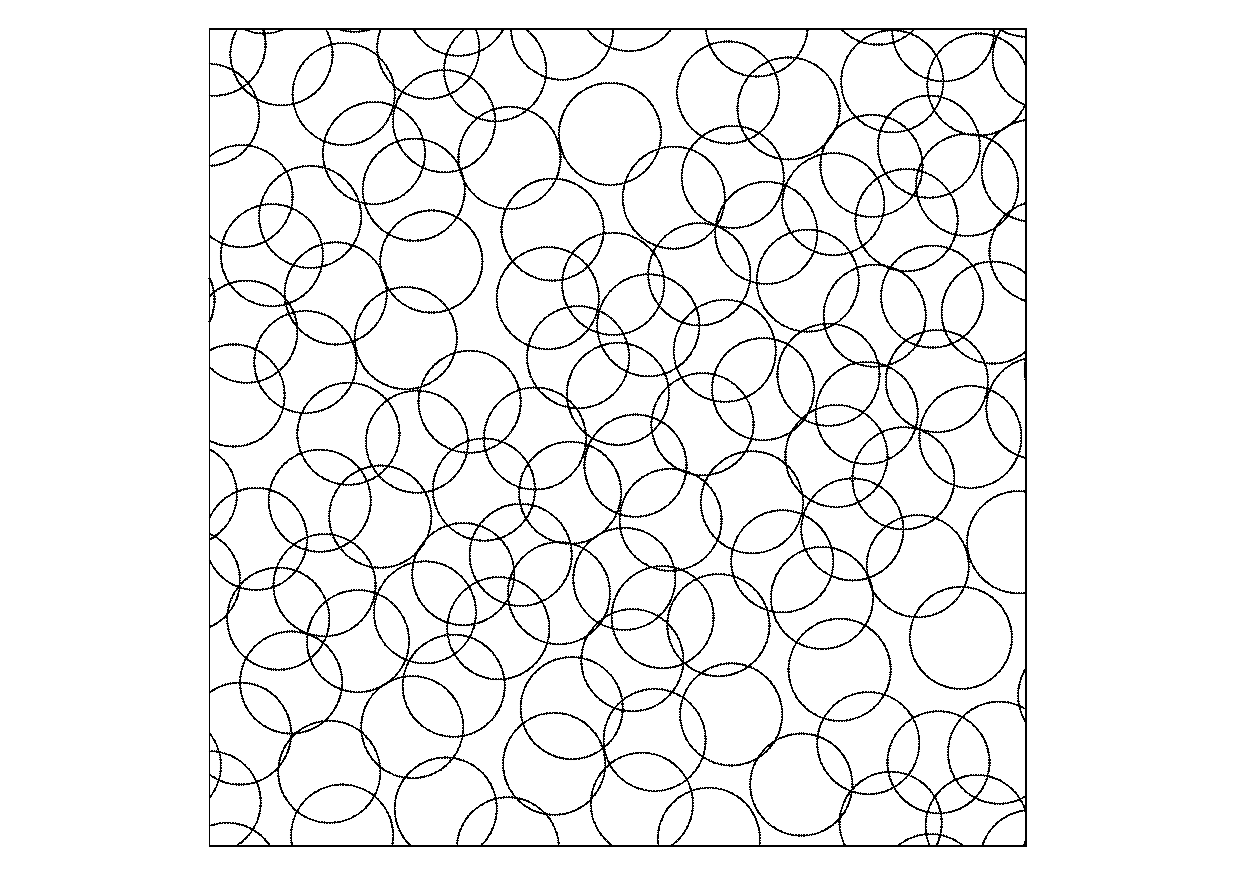}}
  \subfigure[]
  {\includegraphics[width=0.3\linewidth,angle=0]{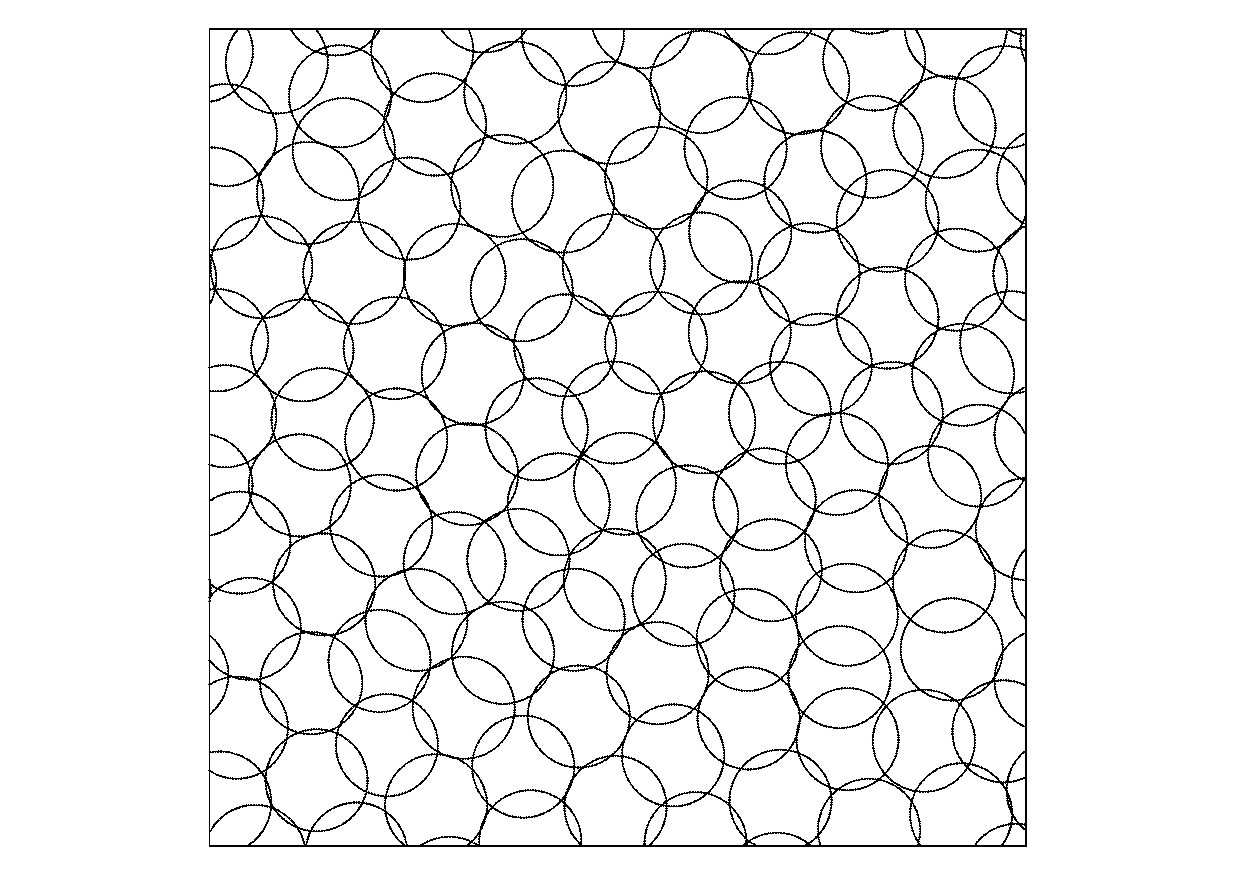}}
  \subfigure[]
  {\includegraphics[width=0.3\linewidth,angle=0]{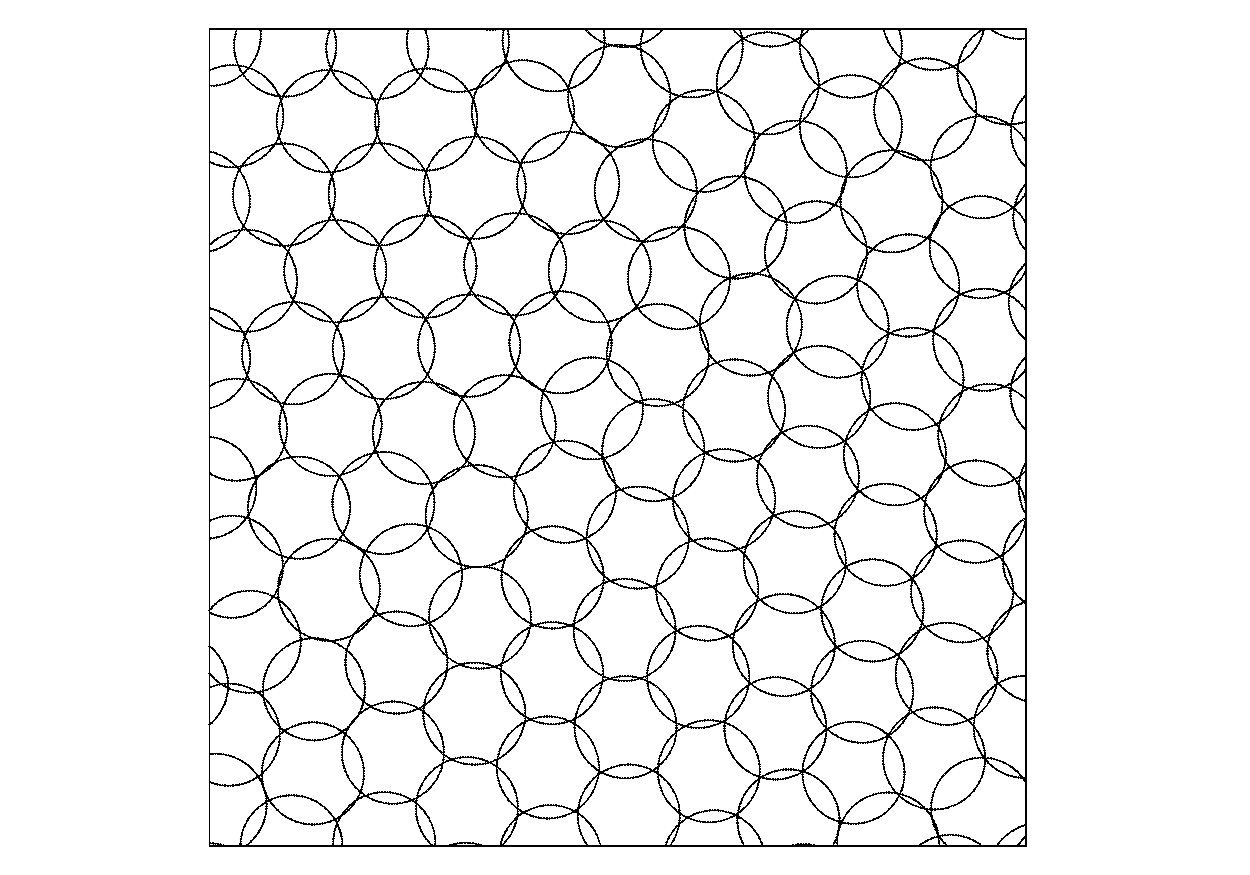}}
\caption{
\label{fig:optimizing}
Illustration of different optimization stages of a $2$-dimensional stochastic
template bank in Euclidean space with a normalized thickness of~$0.38$.
The individual panels are:
(a) Unoptimized template bank that has a covering faction of~$0.89$;
(b) Template bank after $12$ optimization stages (template shifting) that has a covering fraction of~$0.978$;
(c) Template bank after $120$ optimization stages that has a covering fraction of~$0.995$.}
\end{figure*}

\subsection{Performance demonstration}

Figure \ref{fig:optimizing} illustrates the optimization effect 
from barycentric template shifting for a stochastic template 
bank in a 2-dimensional Euclidean space. 
Here, we repeatedly apply the template shifting optimization to the bank.
With an increasing number of such optimization steps
it becomes apparent that the template bank approaches an $A_2^*$ lattice structure.
Since in this simple example the chosen parameter space is quadratic 
with periodic boundary conditions,
a perfect $A_2^*$ cannot be obtained, therefore defects are expected.
This can be avoided by choosing an appropriate size of the parameter space.
Such an example choice for length~$l$ and width~$w$
in 2 dimensions would be $l/w=\sqrt{3}/2$ and a covering radius of the templates 
that is an integer fraction of $w/\sqrt{3}$.

\begin{figure}
\centering
  \includegraphics[width=0.99\linewidth]{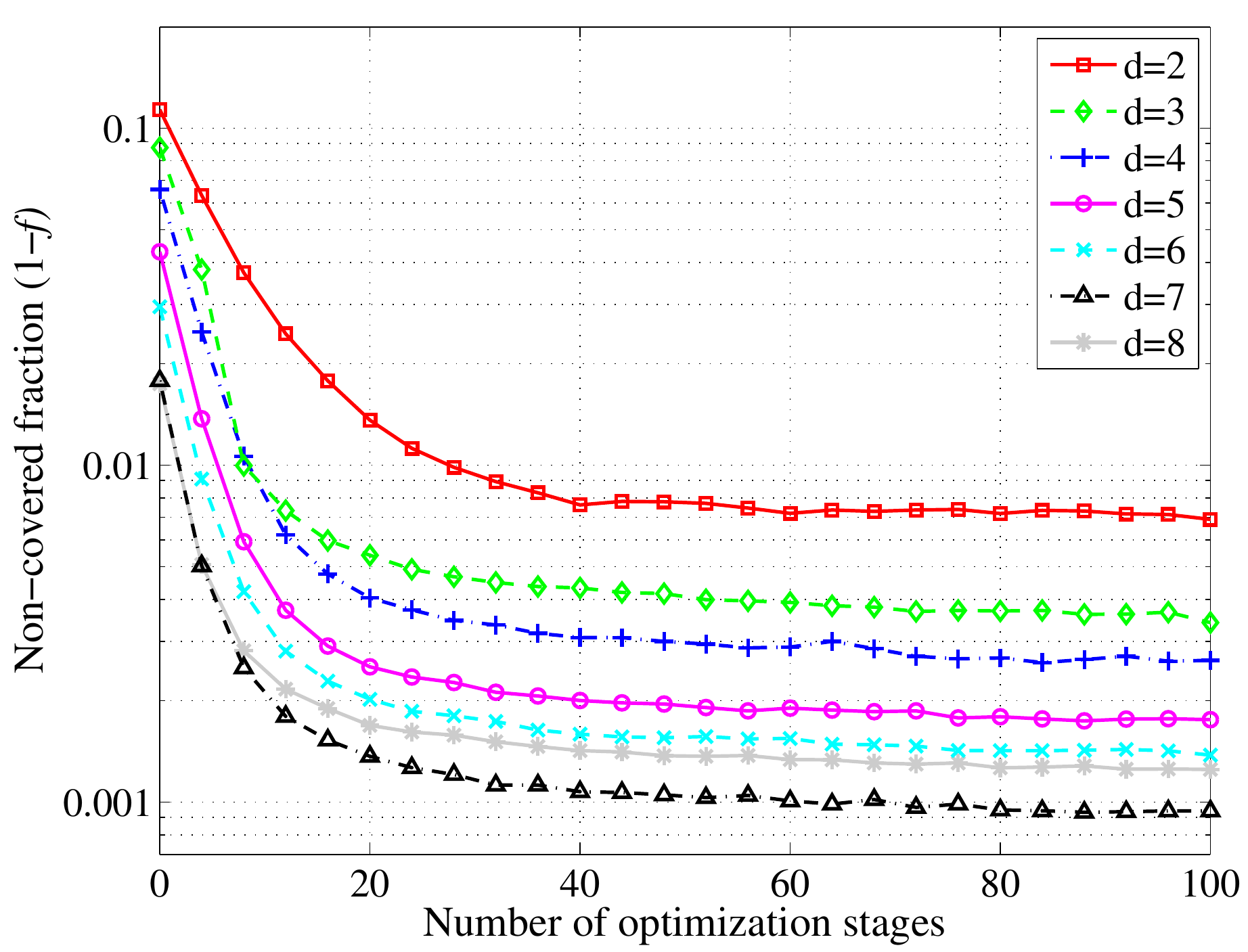}
  \caption{
  \label{fig:optimizing_results}
  The decreasing fraction of non-covered parameter space (that is just \mbox{$1-f$})
  with the number of optimization stages. At each stage, the
  barycentric template shifting method is applied using  maximum 
  shifts of $\epsilon=0.05$. 
  The different curves are for Euclidean spaces~$\mathbb{E}^d$ with
  dimensions from $d=2$ up to $d=8$, using 
  periodic boundary conditions and a template covering radius $r=1$.  
  }
\end{figure}

To evaluate the performance of the template shifting optimization
method, we study the increase in covering fraction~$f$.
Again, for simplicity we consider the Euclidean space~$\mathbb{E}^d$ with up to
$d=8$ dimensions.  For all dimensions, we choose again $r=1$
template shifts are limited to at most $5\%$ of the covering radius, 
so that $\epsilon=0.05$.
The resulting reduction of non-covered parameter space 
(i.e. increasing $f$)
is presented in Fig.~\ref{fig:optimizing_results}.  
As can be seen from the figure, after a few optimization steps
of collective template shifting the non-covered fraction of space
(that is $1-f$) can be significantly reduced. 
Ultimately after a sufficient number of optimization steps the
fraction of non-covered can be decreased by two orders of magnitude
compared to the standard stochastic template bank (corresponding to zero optimization steps).
Recall that this achievement has been made without the addition of any 
extra templates to the bank. Further improvements could eventually
be made by varying or adapting $\epsilon$ during the run time, achieving
a faster convergence or better covering. Similarly, 
replacing the barycentric ``fixed-size'' template shift with some simplex or gradient driven 
downhill method could better take into account the overlapping volume of 
nearby templates and enable even more effective template shifts.

\section{Further examples to test the NCA} \label{s:examples}

For maximum efficiency of the NCA, the cells should be constructed
to adapt to the parameter space structure, e.g. following
the local metric approximation. 
In particular for curved parameter spaces the cell construction
and the determination of neighboring cells requires 
care and can be difficult, in particular in higher dimensions since
the number of neighboring cells grows exponentially with the dimension.  
However, when it is not possible to determine the exact set of
neighboring cells it always is save to just use a somewhat larger set of cells
(that is simpler to determine, but does include cells which are not strict neighbors).
This would only slightly reduce the performance since distances between more
templates have to be computed than actually necessary.  On the other hand,
missing neighboring cells could lead to a more severe issue, since this would
lead to over-covering of templates in the regions of the missed neighboring cells.  
In what follows, we show further exemplary applications of the NCA, 
one related to the choice of coordinates on parameter space,
and one for a parameter space that is curved.

\begin{figure*}
\centering
\hspace{-5mm}
	\subfigure[]
	{\hspace{-5mm}\includegraphics[width=0.33\linewidth,angle=0]{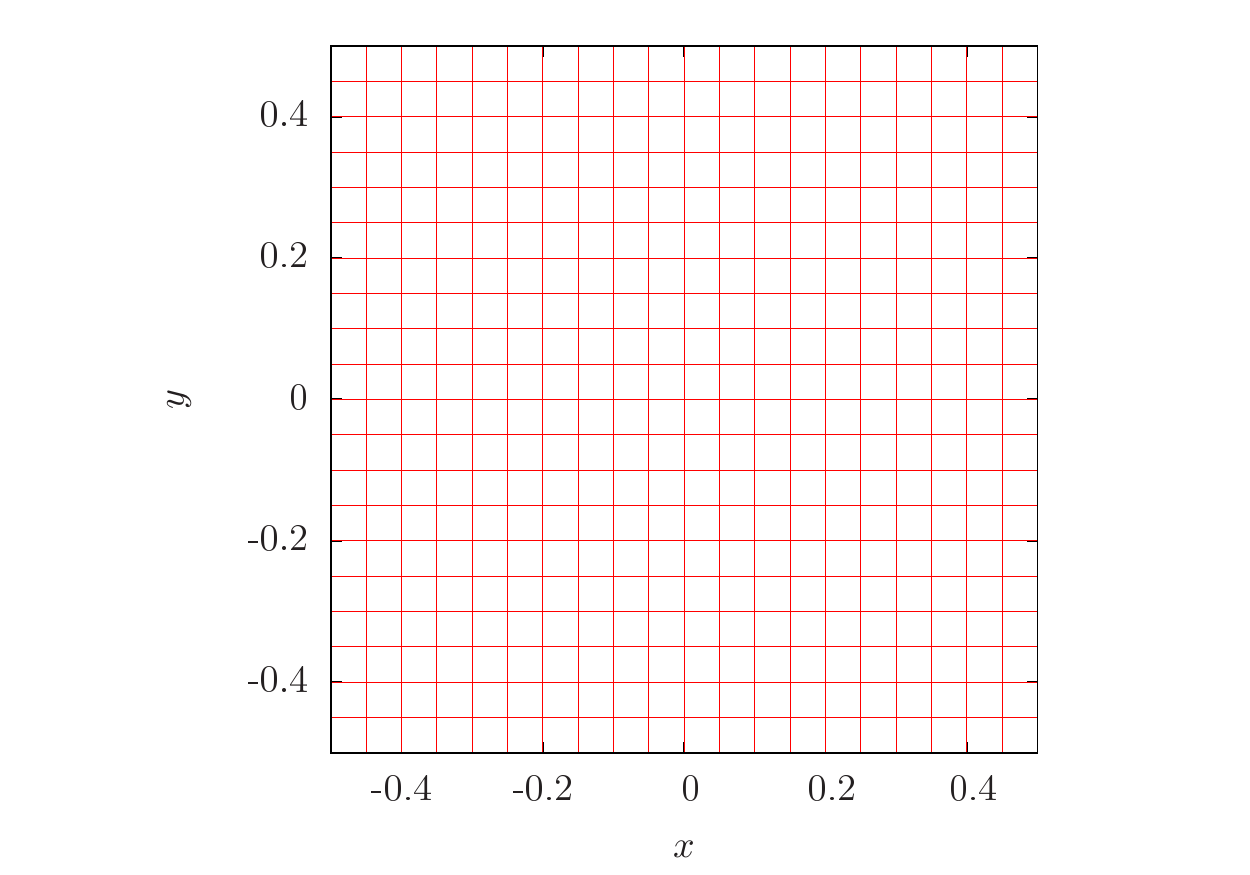}}
	\subfigure[]
  	{\hspace{-5mm}\includegraphics[width=0.33\linewidth,angle=0]{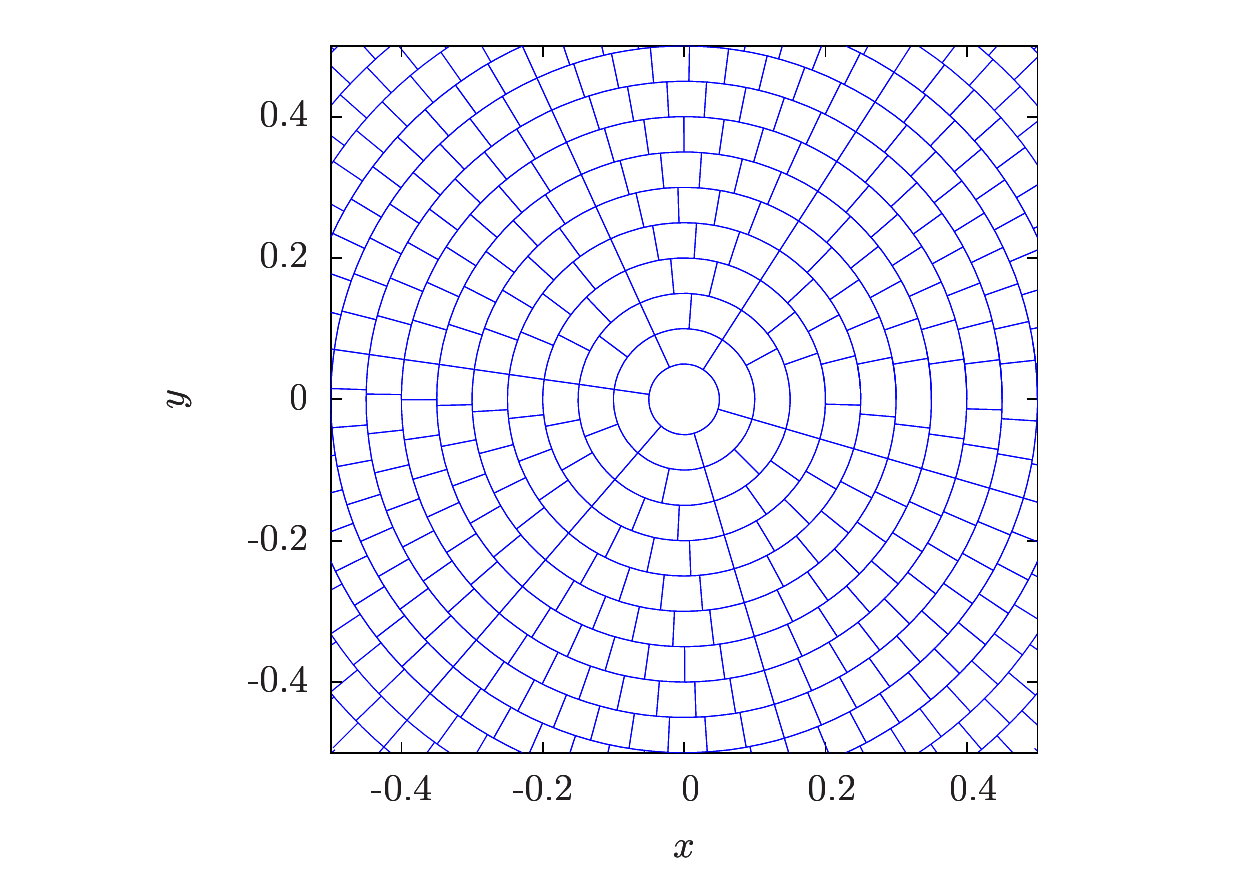}}
	\subfigure[]
  	{\hspace{-5mm}\includegraphics[width=0.33\linewidth,angle=0]{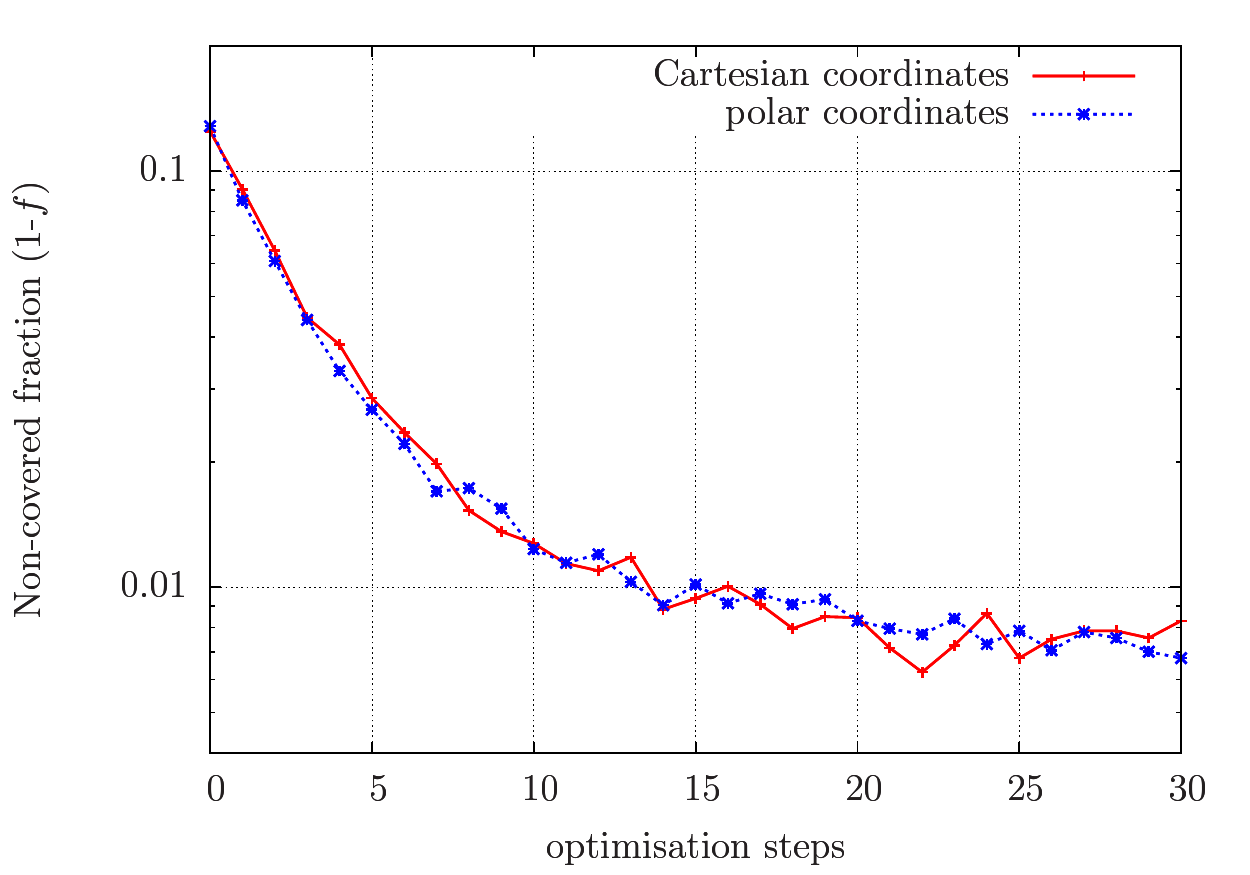}}
\caption{
\label{fig:polar_cartesian}
Example application of the NCA for Cartesian and polar coordinates.
The individual panels are:
(b) Schematic illustration of cell construction when using Cartesian coordinates. 
(b)  Schematic illustration of cell construction when using polar coordinates. 
(c) Comparison of the non-covered fraction $(1-f)$ for Cartesian and polar
coordinates. In this comparison, we employed a template covering radius of
$r=0.01$ and a total number of templates $N=4000$. To determine the
non-covered fraction we used $20000$ Monte Carlo points.
}
\end{figure*}

\subsection{Choice of coordinates}

In principle, the NCA and the optimization are independent 
of the choice of the coordinates. This is demonstrated in the
following example. Figure~\ref{fig:polar_cartesian} illustrates the
parameter space splitting (cell construction) for the NCA
in Cartesian $(x,y)$ and in polar coordinates $(\rho,\phi)$. 
The coordinate transformation is given by $x=\rho \cos \phi$, $y=\rho \sin \phi$
and the distance is computed as $d=\left(x^2+y^2\right)^{1/2}$. 
It is obvious that in polar coordinates the cells are obtained by dividing the parameter space 
into rings of width~$r$, where $r$ is the covering fraction 
of the templates. Each ring is fragmented so that a template covering volume reaches only the
neighboring cells and never the cells beyond.  The neighboring cells are
the adjacent cells in the same ring and any cell in the adjacent
rings which can be ``reached'' by the covering volume of any template in
lying inside the considered cell.  This can also include cells which
have no common boundary points with the considered cell.
Finally, the results of from applying the NCA for both choices of coordinates
are also presented in Fig.~\ref{fig:polar_cartesian}, showing that 
the non-covered fraction as a function of the number optimization stages
(using barycentric shifts, see Sec.~\ref{s:shifts}), 
is effectively the same for both choices of coordinates.

\subsection{Curved parameter space}

\begin{figure*}
\centering
\hspace{-5mm}
	\subfigure[]
	{\includegraphics[width=0.32\linewidth,angle=0]{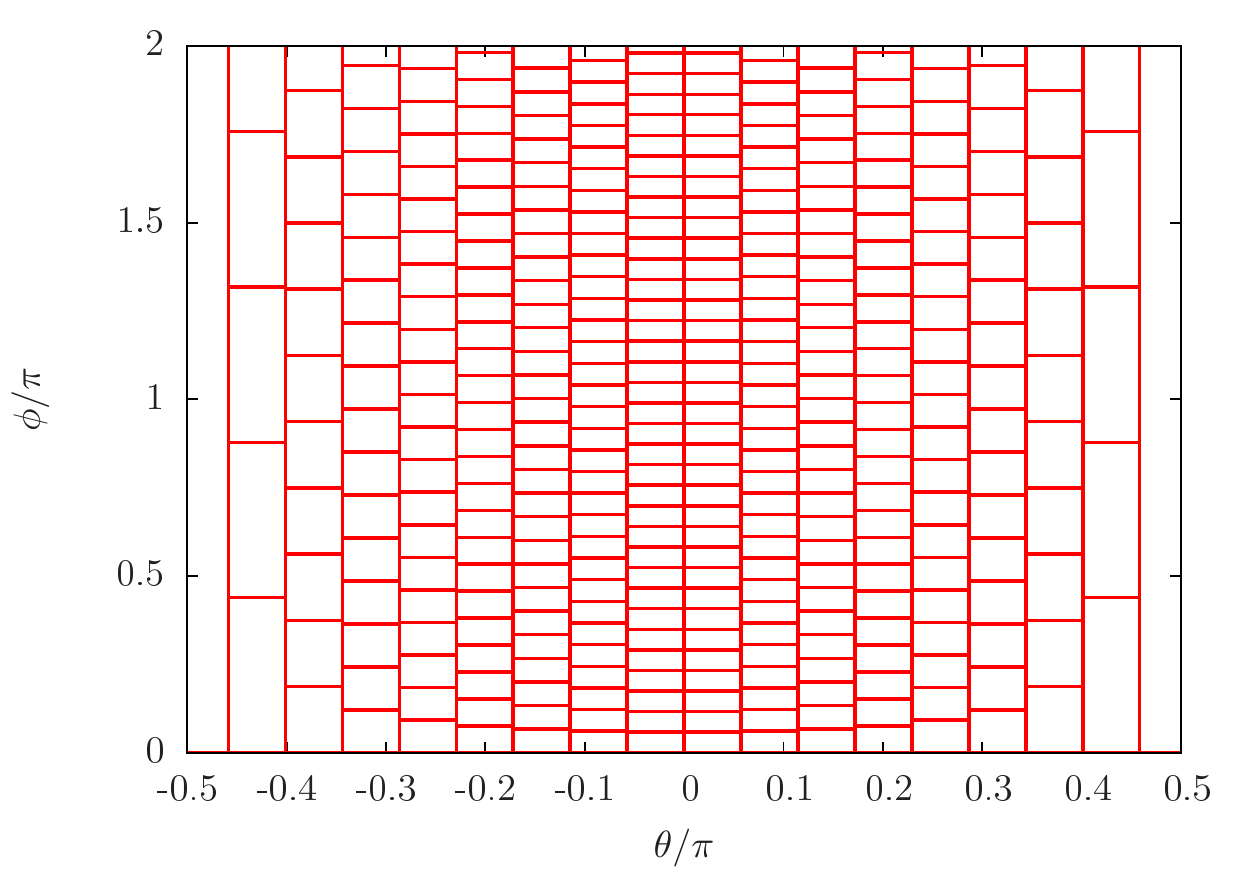}}
	\subfigure[]
  	{\includegraphics[width=0.32\linewidth,angle=0]{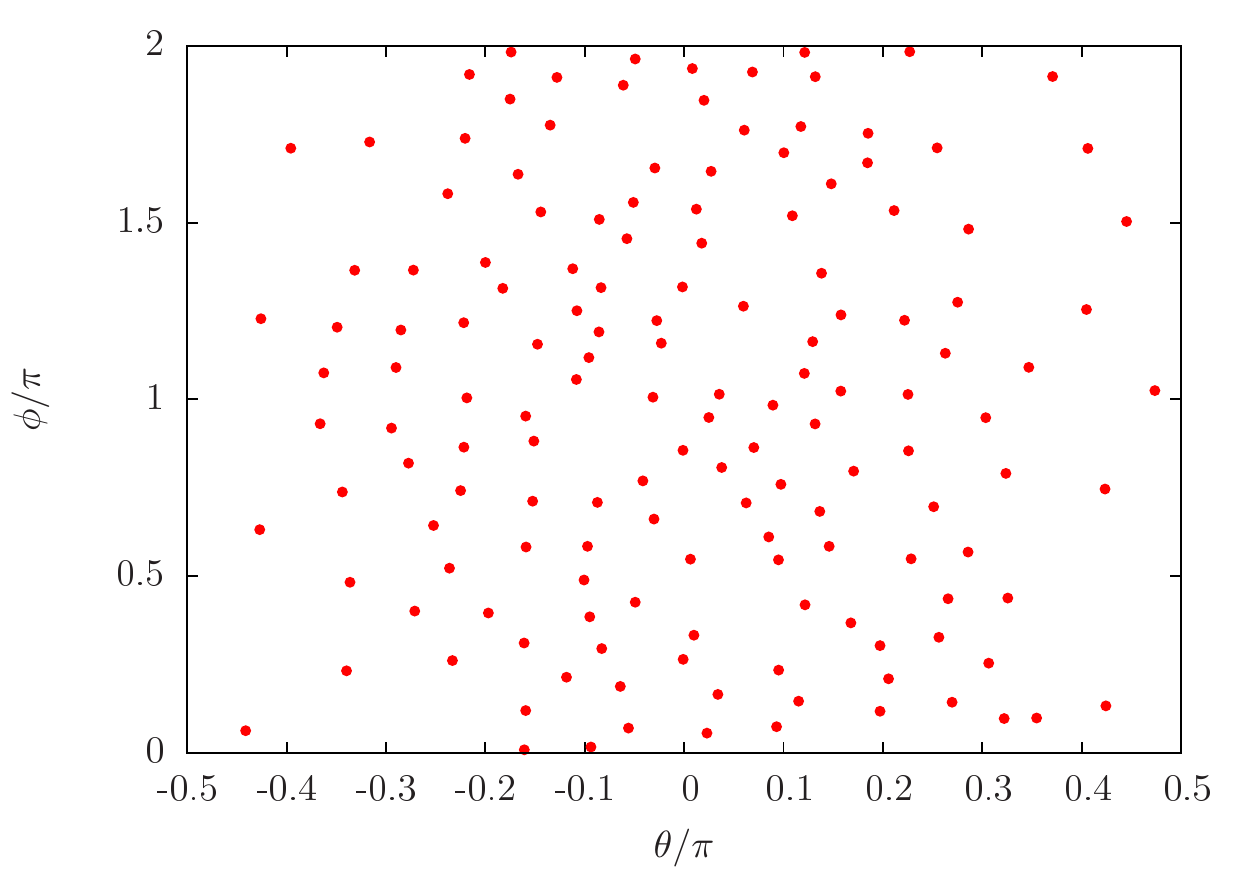}}
	\subfigure[]
  	{\includegraphics[width=0.32\linewidth,angle=0]{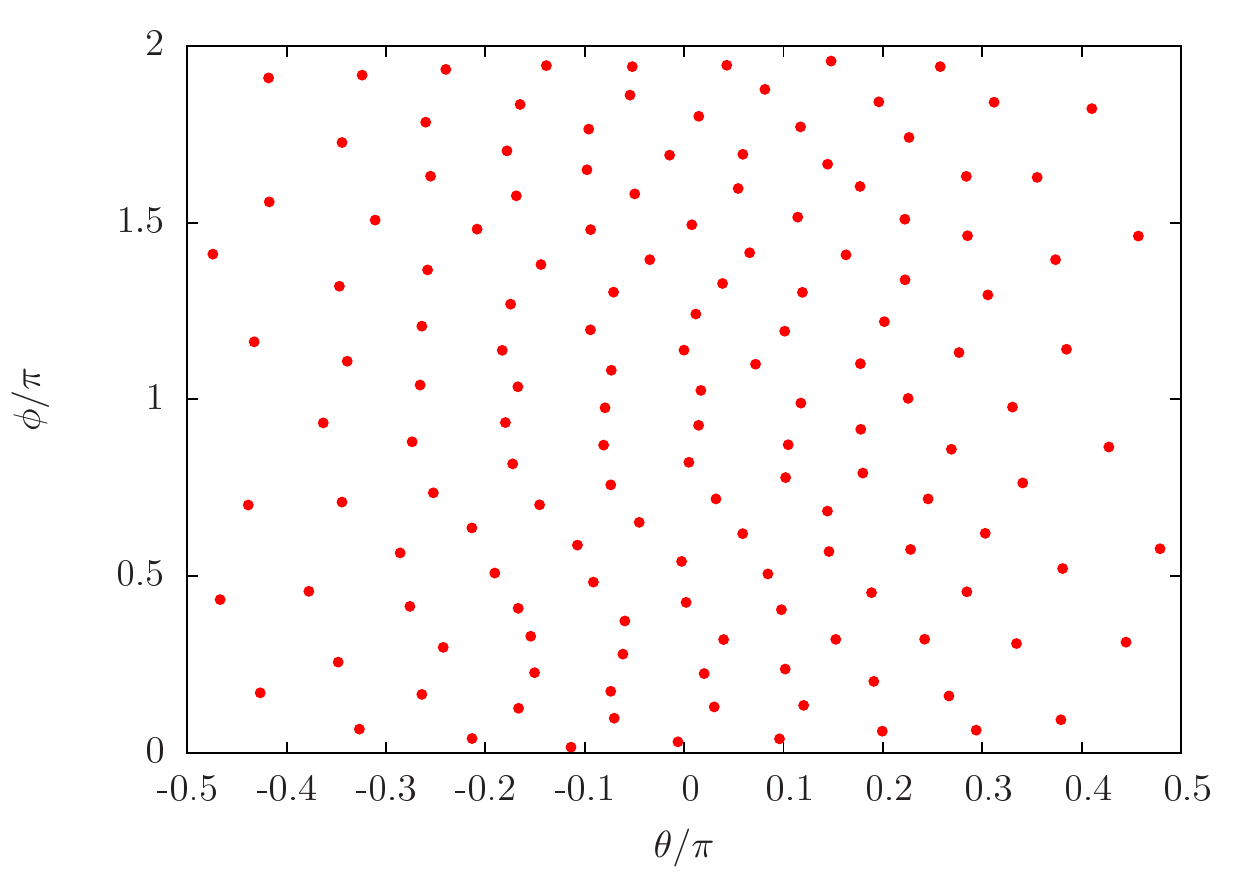}}
\caption{
\label{fig:sphere}
Example application of the NCA to place templates on the sphere,
parameterized by the polar angle $\theta$ and the azimuth angle $\phi$.
Here $\theta=0$ denotes the equatorial plane.
The individual panels are:
(a) Illustration of cell construction in parameter space.
(b) Stochastic template bank from the NCA containing $150$ templates and 
having a covering fraction of $f = 0.858$.
(c) Optimized template bank by the NCA with an improved covering fraction of
$f=0.984$ for the same number of templates. 
The template covering radius has been chosen as $r=0.18$ in this example.
}
\end{figure*}

To illustrate the applicability of the NCA for a curved (i.e. non-flat) parameter space,
we consider generating a stochastic template bank on the sphere -- an example
that was also used in HAS09.
A sphere here means a set of points with the same distance to a center point, 
where unit distance is used for simplicity in the present example. Thus the length element is 
defined as $dl= d\theta + d \phi\cos \theta$ were $ - \pi/2 \le \theta \le \pi/2$ 
and \mbox{$0\le \phi < 2 \pi$}. The cells in parameter space are constructed,
using uniform spacings in the $\theta$-direction. The cell sizes $\Delta \phi$ 
in the $\phi$ direction should depend on $\theta$. Because the cell construction 
should be such that the covering volumes of templates overlap only with 
neighboring cells, we choose $\Delta \phi = r/\cos \theta_b$, 
where $\theta_b$ minimizes $\cos \theta$ within this cell. 
Making $\Delta \phi$ smaller would result in template volumes which 
could reach into non-neighboring cells. In this example, determining
the neighboring cells works similar as described above for polar coordinates. 
For a given $\theta$ one has to find all cells which could have an overlap
with any template inside the considered cell.
In Fig.~\ref{fig:sphere}, the cell construction in parameter space is displayed,
along with the stochastic template generated by the NCA, as well as
the optimized template bank using the barycentric shift method introduced in
Sec.~\ref{s:shifts}.

\section{Generalization of the NCA}
\label{s:generalhca}

In this section, we describe a conceptual idea how to generalize the NCA
for application even to arbitrarily  ``ill-behaved'' parameter spaces.
In general, the smaller the average number of templates per cell the smaller
the number of required distance computations needed by the NCA.
However, for certain parameter spaces, the shape or the size of the template
volumes might be unknown or vary strongly across the space.
This might represent a non-negligible problem in order to meet
requirement~(7) of Sec.~\ref{ss:keyelements},
as one has to choose the size of the cells to be sufficiently large.
It may even lead to extreme situations, where the efficiency gain 
from the NCA can melt away.

\begin{figure*}
   \includegraphics[width=0.8\textwidth,bb=35 100 380 200]{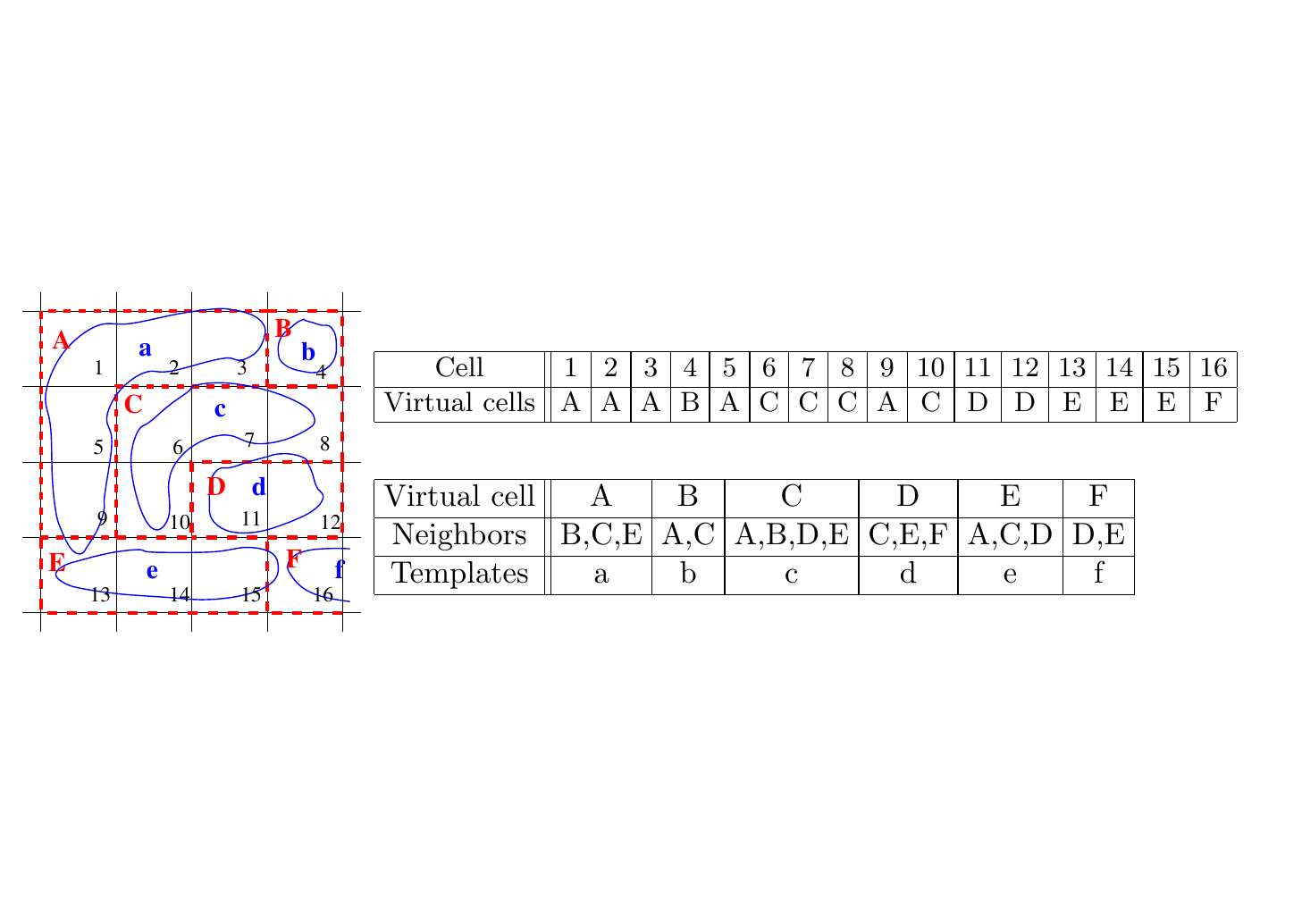}
\caption{
  \label{fig:virtual_hash_cell}
Illustration of NCA generalization by combining cells to form virtual cells
for arbitrary parameter-space structures.
The cell indices are indicated by Arabic numbers.
The virtual-cell indices are labeled with red capital letters.
The templates are denoted by lower-case letters.
The left panel illustrates schematically the borders of the templates, cells, and virtual cells.
The right top table shows a table for the corresponding cell to virtual-cell mapping.
The right bottom table presents a table that lists the neighbors of the virtual cells.
}
\end{figure*}

To address this problem of ill-behaved parameter spaces, we 
suggest the following strategy.
To begin with, set up the cells with a smaller size that actually
violates the requirement~(7) of Sec.~\ref{ss:keyelements}.
Then notice that in some regions of parameter space, a template overlaps
with many cells and not only with neighboring ones. Therefore, one can 
combine these cells to form a single {\it virtual} cell.  Those virtual cells then 
again meet all requirements for the NCA as outlined in Sec.~\ref{ss:keyelements}.  
This basic idea is illustrated in Fig.~\ref{fig:virtual_hash_cell}.
To combine cells the following recipe is proposed:
\begin{itemize}
\item Start with the first cell and place a template inside this
  cell at a random position. Form the first virtual cell labeled~$A$,
  which contains that first cell.
\item Consider the next cell and place a second template inside.
  If the first and second templates are too close to each other,
  the second cell also belongs to the same virtual cell $A$. 
  In this case the second template can be discarded and the
  first template is the representing template for the virtual cell~$A$.  
  On the other hand, if the distance between the two templates is large 
  enough, the second cell forms another virtual cell labeled~$B$, 
  containing the second cell. In this case,
  the first template represents $A$ and the second represents $B$.
\item Continue this scheme subsequently for all other cells and
  test whether the considered cell belongs to one of the existing
  virtual cells. If not, the considered cells forms a new virtual cell.
\item Keep a list in memory which maps all cells to their virtual cells.
\item A virtual cell inherits the neighbors of its containing cells. 
Note that if an inherited neighboring cell is also
a part of the same virtual cell,  this cell has to be removed from
the list of neighbors.
\end{itemize}
This procedure will create a map of virtual cells which cover the 
entire parameter space. While not guaranteed to generate the 
smallest possible virtual cells meeting condition~(7) in Sec.~\ref{ss:keyelements},
this method is a viable solution and more flexible than the basic version of the NCA 
described in Sec.~\ref{s:nca}.
To check whether a specific point in parameter space is covered by
one of the templates, one proceeds as follows:
\begin{itemize}
\item[(1)] Compute the cell index for the parameter-space point.
\item[(2)] Map the cell to the virtual cell and read out the
  indices for the neighbored virtual cells.
\item[(3)] Collect all templates from the template lists of the own and
  the neighbored virtual cells and compute all distances between the
  examined point and these templates.
\item[(4)] If one of the computed distances is smaller than the desired
covering radius, the point is covered.
\end{itemize}
For illustrative purposes, Fig.~\ref{fig:virtual_hash_cell} shows a 
simplified example to which above method is applied.

\section{Conclusions}
\label{s:conclusion}

This paper presents a neighboring cell algorithm (in short NCA) to efficiently 
construct stochastic template banks covering multidimensional parameter spaces.
A core improvement from the NCA is the dramatic reduction in the number
distance computations, achieved by dividing the parameter space into 
separate cells (neighboring cells).
For any point in parameter space we exploit an efficient hashing technique to
obtain the index of the enclosing cell (and thus the parameters of its neighboring templates).
This way, to test if a new candidate template should be added to the bank,
only templates located within the own and neighboring cells have to be considered.
Previous methods~\cite{harry2009,babak_2008} required comparison (i.e. distance computation) 
with all templates already in the bank and thus were considerably more computational 
expensive and eventually prohibitive for large (or also high-dimensional) parameter spaces.
We have demonstrated that compared to the  standard stochastic template bank algorithm,
the NCA can reduce the number of distance computations in a $3$-dimensional Euclidean 
space by about five orders of magnitude. In addition, based on the NCA we have
described a new method to significantly increase the covered 
fraction of parameter space, solely through systematic shifts of the template positions
-- without adding further templates to the bank. 

The NCA is guaranteed to work efficiently if the average number of
templates per neighboring cell is small.  For cases, where the shape and size of the
template volumes vary drastically across parameter space,
this can be eventually become difficult to achieve.  To address this problem, we have presented a 
method to generalize the NCA by combining many neighboring cells to form so-called
virtual neighboring cells. The arrangement of the virtual neighboring cells can 
adapt adequately to the local parameter-space structure (the shape of the 
covered template volumes).

Apart from generating template banks, it should be pointed out that the NCA 
can also be used to efficiently validate the produced bank.
This is usually done by searching synthetic data sets containing simulated signals
and determining the resulting minimum mismatch in each case. The NCA
considerably accelerates this process by avoiding the need of having to the search
the entire template bank for every simulated-signal data set.
Instead, for any given parameter-space position of a simulated signal,
the NCA can readily provide the subset of templates closest to the
signal position, which are the only ones relevant. Templates
further away from the signal location are irrelevant, since those will obviously
have high mismatch with the signal.
Iterating this procedure for a large number of simulated signals
across the parameter space, gives rise to a mismatch histogram to validate
the efficiency of the entire template bank.

The NCA,  including the generalized version presented, 
has applicability in different areas of astronomy.
For example in gravitational-wave searches for inspiral or
continuous-wave sources \cite{Babak+2012,arXiv:1207.7176}, 
exploiting the NCA can potentially offer great efficiency gains.  
In the field of gamma-ray pulsar astronomy, the NCA has already been 
successfully used to construct an optimized stochastic template bank 
to search data from the {\it Fermi} Large Area Telescope 
for a pulsar binary system \cite{Pletsch+2012-J1311}. Further 
details involved and results from these applications of the NCA are 
subject to forthcoming work. 

Directions for a future work also include technical and methodological
improvements. In this work, we implemented the NCA in a parallel algorithm
using {\it OpenMP}\footnote{\url{http://openmp.org/}}
and executed the program on a single system. However, it might be worthwhile
to port to an {\it MPI} version which runs on many compute nodes or use
remote databases to hold the template and the cell table.  The
practicability of such algorithms has to be investigated, particularly since
random access on the entire table ranges is required.
Finally, a further improvement of the optimization method
could be achieved by replacing the barycentric ``fixed-size'' template shift
with some simplex or gradient driven downhill method. This approach would better
take into account the overlapping volume of nearby templates and
enable even more effective template shifts.

\section{Acknowledgments}

We thank Bruce Allen for discussions of the basic ideas behind this work
and for pointing out to us open questions concerning the
convergence behavior of stochastic template banks.
We are grateful for fruitful discussions with
Reinhard Prix, Chris Messenger, Oliver Bock and Carsten Aulbert.
We thank Benjamin Knispel and Gian Mario Manca for discussing
possible further applications and some future directions of this work.
The concept of hash algorithms in numerical physics we owe to
Bogdan Damski.
This work was supported by the Max-Planck-Gesell\-schaft~(MPG),
as well as by the Deutsche Forschungsgemeinschaft~(DFG) through
an Emmy Noether research grant PL~710/1-1 (PI: Holger~J.~Pletsch).

\bibliography{nca}

\end{document}